\documentclass[default,iicol]{sn-jnl}

\usepackage{amsmath}
\usepackage{graphicx}
\usepackage{subcaption} 
\usepackage{comment} 
\usepackage{natbib}

\usepackage{graphicx}%
\usepackage{multirow}%
\usepackage{amsmath,amssymb,amsfonts}%
\usepackage{amsthm}%
\usepackage{mathrsfs}%
\usepackage[title]{appendix}%
\usepackage{xcolor}%
\usepackage{textcomp}%
\usepackage{manyfoot}%
\usepackage{booktabs}%
\usepackage{algorithm}%
\usepackage{algorithmicx}%
\usepackage{algpseudocode}%
\usepackage{listings}%

\begin{document}

\title{Too good to be true: People reject free gifts from robots because they infer bad intentions}


\author*[1]{\fnm{Benjamin} \sur{Lebrun}}\email{benjamin.lebrun@pg.canterbury.ac.nz}

\author[1]{\fnm{Andrew} \sur{Vonasch}}\email{andrew.vonasch@canterbury.ac.nz}

\author[2]{\fnm{Christoph} \sur{Bartneck}}\email{christoph.bartneck@canterbury.ac.nz}

\affil[1]{\orgdiv{School of Psychology, Speech and Hearing}, \orgname{University of Canterbury}, 
\city{Christchurch},
\country{New Zealand}}

\affil[2]{\orgdiv{Department of Computer Science and Software Engineering}, \orgname{University of Canterbury}, 
\city{Christchurch},
\country{New Zealand}}


\abstract{A recent psychology study found that people sometimes reject overly generous offers from people because they imagine hidden “phantom costs” must be part of the transaction. Phantom costs occur when a person seems overly generous for no apparent reason. This study aims to explore whether people can imagine phantom costs when interacting with a robot.
To this end, screen or physically embodied agents (human or robot) offered to people either a cookie or a cookie + \$2. Participants were then asked to make a choice whether they would accept or decline the offer.
Results showed that people did perceive phantom costs in the offer + \$2 conditions when interacting with a human, but also with a robot, across both embodiment levels, leading to the characteristic behavioral effect that offering more money made people less likely to accept the offer. While people were more likely to accept offers from a robot than from a human, people more often accepted offers from humans when they were physically compared to screen embodied but were equally likely to accept the offer from a robot whether it was screen or physically embodied.
This suggests that people can treat robots (and humans) as social agents with hidden intentions and knowledge, and that this influences their behavior toward them. This provides not only new insights on how people make decisions when interacting with a robot but also how robot embodiment impacts HRI research.}




\maketitle







\section{Introduction}
Humans are social agents who who engage in social cognition to help them interacting with other humans and predicting other's behaviors, such as attributing them a mind and intentions (see \citep{Aronson2018, Dennett1987, Gray2007}) such as selfishness. People expect others to be selfish, at least to some extent \citep{Miller1999}. However, unlike selfishness, previous research has shown that people who are excessively generous for no known reason lead others to imagine hidden reasons to justify such generosity \citep{vonasch2022, HOSEVonasch}. While conventional economic decision-making models would suggest that people are more keen to accept an offer when more money is proposed (see ``price effect'' in \citep{Frey1997}), recent research showed that you might consider this offer suspicious and be less likely to accept the offer \citep{vonasch2022}. To consider behaviors as suspicious, people need to first consider the overly generous person as a social actor but also to imagine what his or her intentions are. It is not known however, how humans would react if the agent making an overly generous offer was not a human but a robot.

While it has been shown that people can attribute intentions to social robots \citep{Marchesi2019, Thellman2017}, robots are still different from humans, physically and behaviorally. Robots are expected to interact with humans and help them with different tasks in different contexts, such as healthcare services. It is important to know how robots need to sufficiently justify their behaviors to increase the likelihood of people accepting their help. Exploring people's behaviors in economic situations involving a robot making an outstanding offer would provide this information by giving insights concerning the manner people perceive and behave with robots but also how those robots should interact with humans to get adequate interaction.




\subsection{Heuristic of Sufficient Explanation Model}
People use explanations in order to understand the world (e.g., nature, people, mental states, behaviors, etc). When communicating, people are vigilant about information provided by others and tend to determine whether the information they received is reliable in order to avoid deception and/or misinformation \citep{Sperber2010}. Little is known however about how people act when no sufficient information about specific deeds has been provided. When someone is acting in a way that is not expected by other people, such as being overly generous, people try to understand the reason why this person is acting that way \citep{Ratner2001}.

While classic economics models suggest that people are mainly rational and aim to maximize gains \citep{Frey1997}, other scholars found that some people show a loss aversion bias \citep{Kahneman1979}. For them, avoiding losses is more important than prioritazing gains. In this context, what would happen is someone is overly generous with someone by giving more than what would already be expected? \citet{vonasch2022} looked into this matter and found that when someone is making an offer that seems ``too good to be true'', people are more likely to refuse it. Refusing generous offers seems to go against common sense and rationality as we may expect people to accept generous offers if no explicit reason is given to doubt the offer.

Vonasch and collaborators described the tendency of money backfire in their Heuristic of Sufficient Explanation model -- or HOSE -- \citep{vonasch2022, HOSEVonasch}. This model aims to explain people's decisions based on the explanations they received about the situation. More specifically, it proposes that that when an explanation about an event seems insufficient, people will look for hidden reasons to explain why this happens to fill the gap, such as trying to find the ``real'' reasons and intentions of the person making the offer. In the case of over generous offers, people will not necessarily choose to maximize their monetary incentives but can rather make their decisions based on underlying beliefs and expectations. This model is applicable in different contexts such as economic decision-making, belief in conspiracy theories, or judgements of other agents' bad intentions \citep{HOSEVonasch}.

This model highlights that people are suspicious when explanations are considered as insufficient. This suspicion serves to protect themselves from danger and understand and predict future behaviors better, people will look for reasons that may be correct or incorrect. If imagined reasons are correct, people will have behaviors adapted to the situation. If reasons imagined were to be misleading, people will imagine phantom costs that may negatively influence their behaviors.

\subsection{Phantom Costs}
The HOSE model explains that when people receive insufficient explanations, they imagine hidden reasons to fill the gap. These hidden reasons are often phantom costs.  People imagine phantom costs when someone makes and overly generous offer without explicit reasons to justify why \citet{vonasch2022}. This is grounded on the idea that people are social agents, not only self-interested economic agents. For phantom costs to occur, three criteria need to be met \citep{vonasch2022}. First, people expect others to follow the norm of self-interest. Second, this expectation has to be violated. Third, no sufficient reason is provided to justify this violation. In the face of such unselfish behavior --if no sufficient reasons justify this--, people start being suspicious and looking for reasons explaining unjustified behaviors \citep{Ratner2001}. Because people generally expect other humans to pursue self-interest, they imagine phantom costs when a human offers too much in a transaction. 

\citet{vonasch2022} showed that people imagine phantom costs when a person was offering people money to eat a cookie. Why would this person not want them for him/herself? Moreover, why would this person pay others \$2 to eat them. This behaviour is in conflict with the self-interest expectation. Furthermore, he/she did not offer any explanation for his generosity. All three criteria described earlier for phantom costs to occur are met and hence, people are likely to reject the offer. This shows that phantom costs lead to money backfire where offering money mitigates people's acceptance rate in the offer. This money backfire effect is explained via the reduction of the net psychological value, including both the money and phantom costs associated to the transaction \citep{vonasch2022}. This is because money modifies the potential costs people might perceive during a transaction. These phantom costs have been shown not only with food (cookie experiment) in several contexts such as buying cheap flights, being paid to take a specific path etc \citep{vonasch2022}. As a baseline for our study, we aim to replicate the results of the experiment involving cookies \citep{vonasch2022} and hence lead us to the first research question:

\begin{enumerate}
    \item[] \textbf{RQ1:} Do phantom costs exist in a human-human interaction? - This is a replication of \citet{vonasch2022}.
\end{enumerate}
    

We consider that these phantom costs do occur because people infer intentions, mostly bad intentions, to those making the offer (e.g., the person poisoned the cookies). But what about robots? People may or may not expect them to act like other social agents and pursue their own self-interest.
The present research tested whether people ascribe such intentions to social robots.

\subsection{Intention Attribution}
Intentional stance corresponds to predicting and explaining other's behaviors by attributing them mental states such as intentions and beliefs \citep{Dennett1987}.
Ascribing intentions (in the sense of having a mind) to robots seems twofold. One might argue that attributing their owns intentions to robots is not possible since they are mere machines and should not be treated as intentional agents.
On the other way, some theories, such as the \textit{Media Equation} and the \textit{Computer as Social Actors} theories \citep{Nass1994, ReevesNass1996}, argue that people behave with technology, such as media or computers, as they would do with another human. If so, if people do ascribe intentional behaviors to robots, it is likely that these behaviors can be perceived as suspicious and leading to phantom costs.

Some scholars focus their research on the ascription of intentionality to robots  and consider that people can adopt the intentional stance towards robots, at least to some extent \citep{Marchesi2019}. \citet{Roselli2023} showed education background differences where people who have a high demand of mentalizing skills (e.g., psychotherapists) do attribute intentionality to robots more often than those who do not need these skills (e.g., roboticists). Another study was conducted by \citet{Thellman2017} where participants had to assess screen embodied agents that were a human and a robot doing different behavior on different criteria including the attribution of intentionality. Overall, similar results were found between both human and robot displayed behaviors. \citet{deGraaf2019} extended this idea by analyzing the explanations of both humans' and robots' behaviors. To do so, they presented verbal descriptions of robots' behaviors (so without visual representation of the robot) and participants had to explain in their own words why the robot was performing this action. They found that people use similar explanations for both agents, meaning that people do attribute intentionality to robots.

According to the aforementioned theories and claims, we expect phantom costs to occur not only in a human-human interaction but also in a human-robot interaction.
Phantom costs have the capacity to influence human decision-making processes and the way they will perceive their interlocutor.
Understanding the relevance of phantom costs in HRI is crucial. This study aims to provide new insights into how humans make decisions when they interact either with a human and a social robot. Phantom costs are not directly visible to people and can even just be imagined if perceived insufficient information is provided. Therefore, they are challenging to predict. Being aware of their existence in HRI will help to better understand hidden and contextual factors influencing human decision making in HRI. While several studies explore decision making in the context of human-robot interaction, none, to our knowledge, explored such hidden factors that influence human decision-making. A need for transparency is necessary. Transparency is defined in \citet{Wortham2017} as ``the extent to which the robot’s ability, intent, and situational constraints are understood by users.'' In this context two research questions emerge:

\begin{enumerate}
    \item[] \textbf{RQ2:} Do phantom costs exist in a human-robot interaction?
    \item[] \textbf{RQ3:} Is there a difference in terms of phantom costs when the agent is a human versus a robot?
\end{enumerate}

Whereas \citet{vonasch2022} highlighted phantom costs in online and in-person studies, results seemed to be heading in the same direction but with different amplitudes. The authors showed, for the in person and online experiments respectively, an acceptance rate of 39.4\% and 55.2\% when the offer was not associated with a monetary outcome compared to 20.5\% and 34.6\% when it was the case. However, even if these both study designs were a bit different, differences between these settings and more specifically in HRI (after having discusses about \cite{Thellman2017} above) need to be discussed further as both study settings and robot embodiment are known to influence results in HRI.

\subsection{Online and In-Person Studies}
Recruiting participants using crowdsourcing platforms, such as Prolific or Amazon Mechanical Turk, is faster, cheaper, and allows a higher demographic diversity of participants \citep{Buhrmester2011}. More importantly, online studies ensure that all participants will have exactly the same interactions allowing better control \citep{Naglieri2004}. However, using these online services might provide different conclusions as participants are not in a real situation. 
Facing this debate, scholars attempted to determine if results are similar between online and in-person settings. While some considered that results are pretty similar \citep{Gamblin17}, some others discussed potential risks of using online settings such as the deterioration in results over time \citep{Tang2022} or the use of sophisticated bots \citep{Griffin2022}. As specified earlier, \citet{vonasch2022} found similar conclusions in both settings. It therefore seems important that we validate our results in terms of robot embodiment, not physically embodied in the online setting, and physically embodied in the in-person setting. This will show whether robot embodiment influences the perception of phantom costs.

Unfortunately, in the field of human-robot interaction, not all laboratories and scholars have a real robot, questioning the usability and similarity of results between according to agent's embodiment needs to be addressed.

\subsection{Robot Embodiment}
\label{media}
Robot embodiment is an important debate in HRI research.
Scenario-based media are frequently used in the literature in place of real interactions between a human and a robot. They are defined as ``discounted media'' as only partial information is provided \citep{Randall2023, Xu2015}. For the sake of clarity, we will refer to the discounted media as ``representations.'' These robot representations could be of different types such as pictures, videos, virtual agents, or simply text-written scenarios.
\citet{Xu2015} used the model of prototyping fidelity \citep{McCurdy2006} to distinguish these representations according to their fidelity and interactivity. The former corresponds to the ``level of detail a feature or sequence is represented'' (e.g., visual, tactile). The authors also defined interactivity as ``the level of feedback that a product or a system affords an evaluator’s engagement with the product.'' The authors gave the example of high interactivity when the user can directly interact with the robot without a limitation, and a low level of interactivity happens when the user is a mere observer of the interaction.

While a live interaction is high for both fidelity and interactivity levels, text and pictures are low. They describe the Wizard-of-Oz method (i.e., controlling the robot without people to know) as high in terms of interactivity but intermediate in terms of fidelity. Even so, it is already known that pictures can elicit human reactions and emotions, the use of the Self-Manikin Assessment associated with the International Affective Picture System is a good example \citep{IAPS2017}.

Several scholars explored the effect of robot embodiment. 
Screen embodied robots have several advantages over physically embodied agents. They allow researchers to run studies online, and no risk of having the robot crashing during an experiment can happen \citep{Belpaeme2020}. Using such screen embodied robots is nowadays easy since \citet{Phillips2018} who created a database of real robots photos. \citet{xu2012} specified that text-written scenario proposes high flexibility in their creation but also a reasonable power of the results on user acceptance.

It is of course possible to go beyond images. In their review, \citet{Jung2021} found that about 10\% of the HRI research evaluating the usability of social robots are using photos, and another 10\% are using videos. We can expect these percentages to be extended and more or less similar to all the HRI research. It seems of the utmost importance to know if results are the same using different level of embodiment (screen and physically embodied). \citet{Woods2006} found moderate to high agreement on several between videos and live interactions on preferences for specific robot behaviors (i.e., the manner the robot was approaching the human). While photos are static representations of the robot, videos can show the dynamic of the interaction with a more realistic viewpoint of the robot \citep{xu2012}. However, our study did not focus on this dynamic but more on the perception of the robot. Moreover, looking at the crowdsourcing platform Prolific, we can easily see that studies which require participants to use sounds (so most likely to be a video) have a very slow recruitment rate. At the day this paper is written, two out of the three proposed studies require sound and are recruiting their participants since weeks and even a few months. As an example, a last check on Prolific on January 29, one of the study required 800 participants. For now, 614 people participated in the study, and the researcher recruited only 5 participants the last two days.They seem to not get their required sample size easily. It seems reasonable to think that requiring sound mitigates the likelihood of recruiting participants and therefore decreases the main advantage of the online studies which is to recruit faster. For our purpose, it seems reasonable to think that a photo of the agents is enough.

While some scholars showed more positive interactions in favor of physical robots, such as greater trust \citep{Bainbridge2011}, others found better results for virtual social robots or even mixed effects (see \citet{Oliveira20} for a review). \citet{Li2015} also analyzed 39 effects reported in HRI papers and showed that 79\% were in favor of physically present robots and only 10\% were in favor of non physically present robots. \citep{Thellman16} proposed that it is not really the physical presence that matters but rather its social presence. The latter is defined as experiencing artificial objects as social actors as they manifest humaneness, such as a conversation \citep{Lee2004}. This occurs when people respond to agents as if they were humans \citep{Lee2006}. We consider that both plays a role in the attitudes towards robots but that the one does not replace the other. If the social presence seems to be sufficient, scenario-based media should prove to be enough.

\citet{xu2012} highlighted that different scenario representations triggers different human behaviors. More specifically, live scenario were superior in several aspects they have measures but text and videos provided pretty good results when it was about general attitudes towards robots.

While scholars do not seem to agree on the efficiency of using scenario representations in HRI, it is still unclear if different robot representations can replace in-person human-robot interactions. It seems therefore important to conduct our study using two types of embodiment, on a screen and physically present, to contribute to the collective understanding of the issue and understand better how scenario representations influence phantom cost perceptions. If similar results were found between scenario representations and real life human-robot interaction, online studies might be conducted by researchers, in this context of suspicious behaviors and phantom costs, saving money and time. Criteria people use to make their decisions should also be collected to understand better how they make their decisions. This information lead us to the three last research questions of this paper:

\begin{enumerate}
    \item[] \textbf{RQ4:} Are there interaction effects between the Agent, Offer and Embodiment?
    \item[] \textbf{RQ5:} What are the criteria participants use to make their decision?
    \item[] \textbf{RQ6:}  Is there a difference in terms of phantom costs when the agent is physically or screen embodied?
\end{enumerate}

\subsection{Objectives}
This study is a partial replication of the work of \citet{vonasch2022}, enabling us to verify our method by comparing the results. We extend it by the perception of phantom costs in the context of human-robot interaction to explore differences in terms of Agent and their Embodiment.
Additional exploratory research question will aim to know if there are phantom costs perception differences between genders and what are the criteria people use to make their decisions.

Exploring the perception of phantom costs in HRI will allow us to know whether a broader context is taken into account by people when interacting with a robot to make decisions related to the interaction. If so, phantom costs can occur when explanations seem insufficient and hence scholars should keep this in mind when conducting HRI studies.

\section{Method}
We conducted a between-participants design using three factors. An \textit{Agent}, either a human or a robot, was making an \textit{Offer} that was either a cookie or a cookie associated with \$2. The Agent was either physically or non-physically embodied. While for the physically embodied agents, the study was conducted in-person, the study for the non-physically embodied agents was conducted online. To conduct the latter, people saw a picture of the real agents on their screen. For the sake of clarity, we will refer to these two conditions as \textit{physical} and \textit{screen} embodiment. While it is only one study with three factors, we chose to describe the method of both embodiment conditions into two different sections.

\subsection{Part A: Screen Embodiment}
We conducted a between-participants design where a screen embodied agent, either a human or a robot, was making an offer that was either a cookie or a cookie associated with \$2. The pre-registration for this study is available at \url{https://aspredicted.org/XRZ_ZH7}. Initially, a first version of the screen embodiment study has been conducted only using the screener ``fluent in English'' but it seems to not have be sufficiently controlled because of the large demographic variety (people from 27 different countries participated in the study). Some participant seem to not have correctly understood the scenario or even answered in Spanish. Also, the point of this study was not to create specific groups to explore potential cultural differences in the perception of phantom costs. While we suspect it may exist, we have chosen to rerun the study using more specific screeners via Prolific. For the sake of clarity and transparency, please find analyses included the first version in \autoref{appendix:a}. Note that the conclusions drawn are alike.

\subsubsection{Participants}
481 participants were recruited via the Prolific platform\footnote{\url{https://www.prolific.com}} to take part in this online study. Inclusion criteria were to be located in the USA, have the American nationality, and fluent in English.

\subsubsection{Measurements}
Two independent variables were used for this study: Agent and Offer. The Agent was either a Human (coded as 0) or a Robot (coded as 1). The Offer was either a Cookie (coded as 0) or a Cookie + \$2 (coded as 1). Participants were asked to say if they would accept or refuse the offer using a forced binary choice (coded as 0 if they refused or 1 if they accepted). On a new page, they had to rate statements on 7-point Likert scales. They were then asked to explain their choice in a text-entry box and provide some demographic information (i.e., gender, age, Hispanic origins, and race).

\subsubsection{Material and Place of Study}
\begin{figure*}[htbp]
  \centering
  \begin{subfigure}[b]{0.45\textwidth}
    \includegraphics[width=\textwidth]{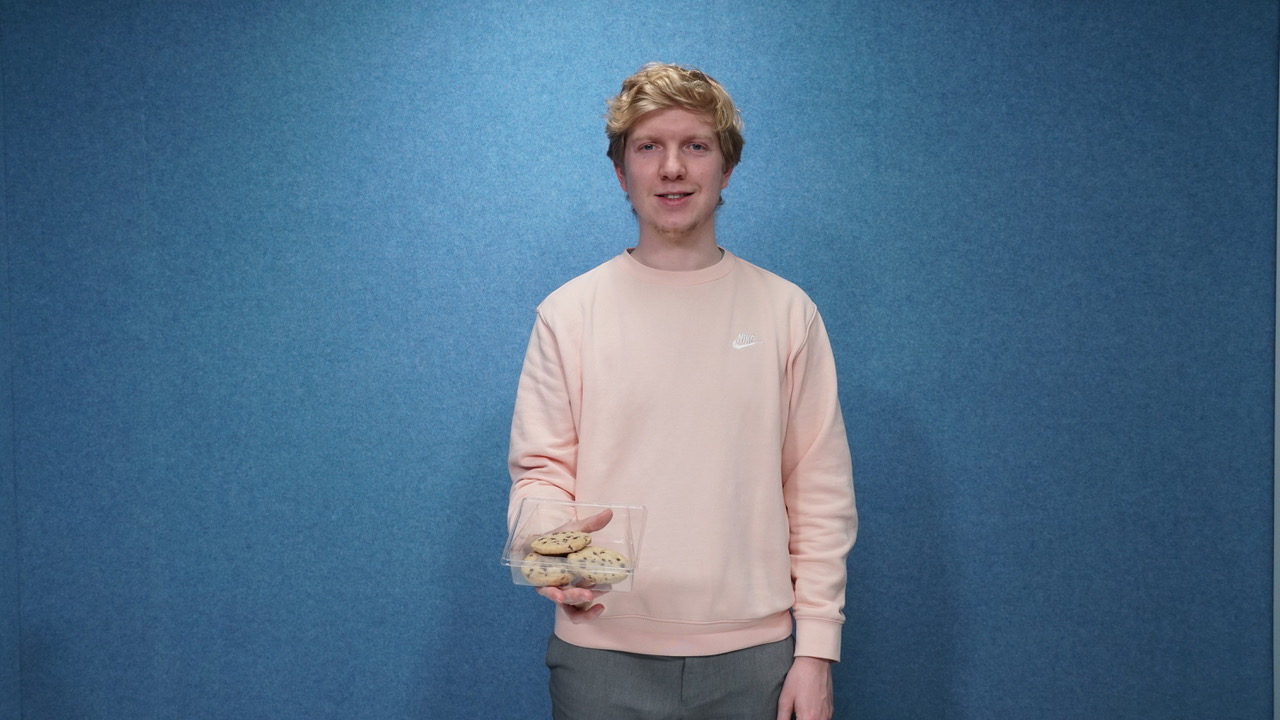}
    \caption{Human - Cookie}
    \label{fig:sub1}
  \end{subfigure}
  \hfill
  \begin{subfigure}[b]{0.45\textwidth}
    \includegraphics[width=\textwidth]{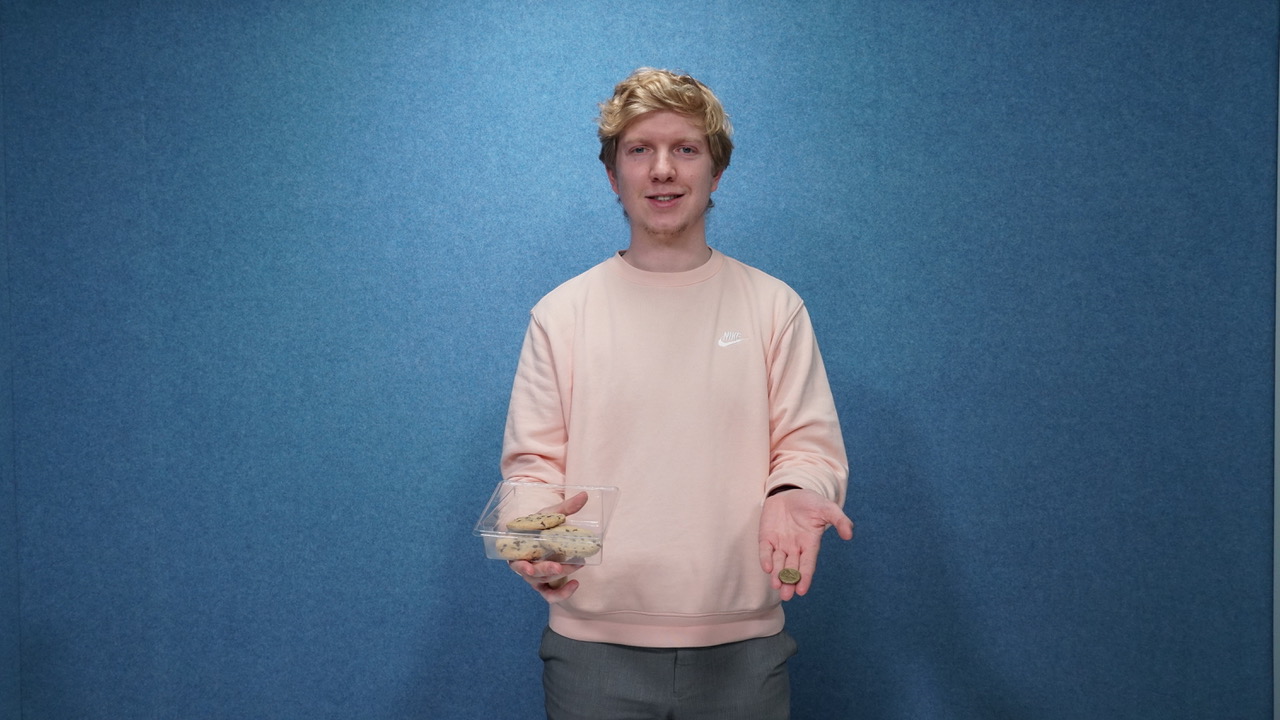}
    \caption{Human - Cookie + \$2}
    \label{fig:sub2}
  \end{subfigure}

  \begin{subfigure}[b]{0.45\textwidth}
    \includegraphics[width=\textwidth]{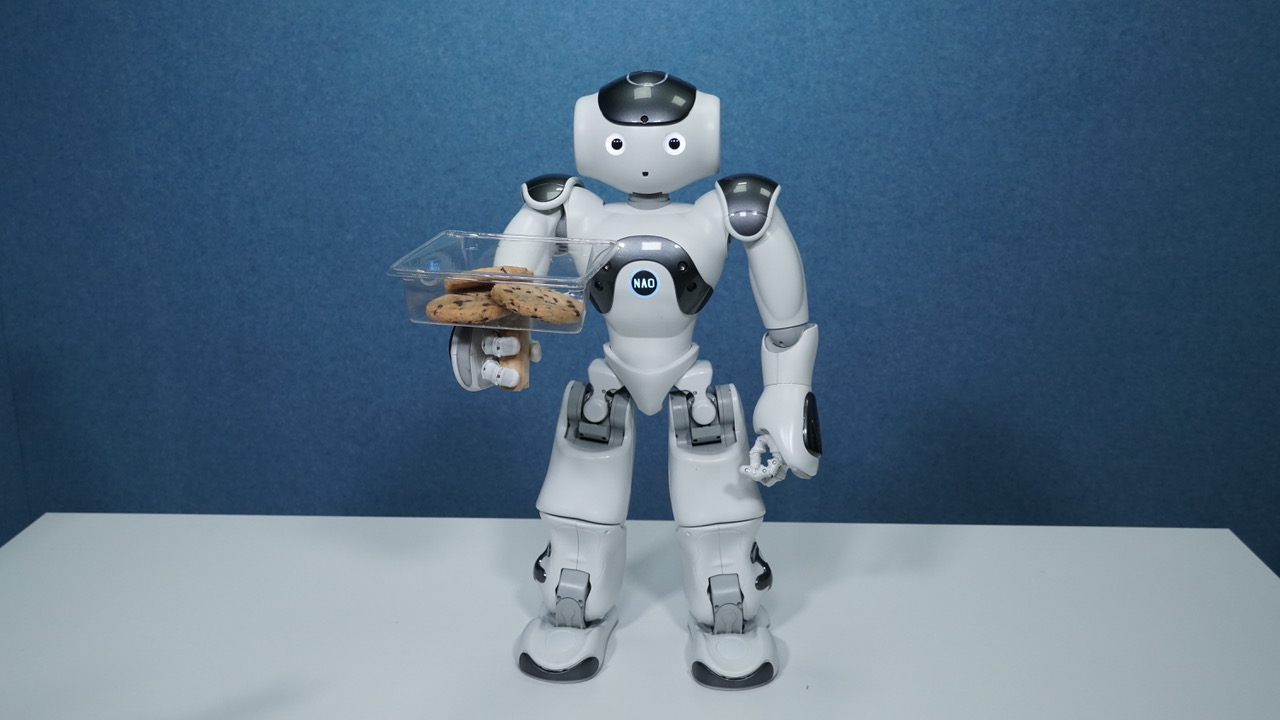}
    \caption{Robot - Cookie}
    \label{fig:sub3}
  \end{subfigure}
  \hfill
  \begin{subfigure}[b]{0.45\textwidth}
    \includegraphics[width=\textwidth]{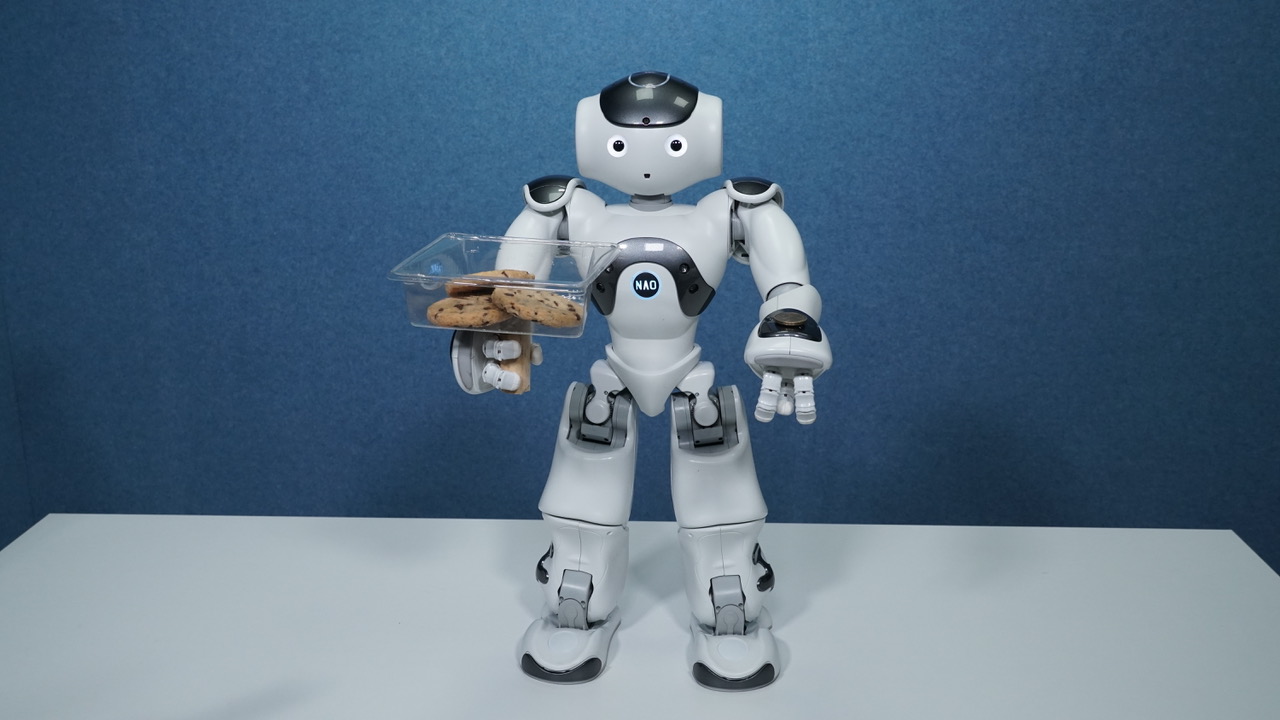}
    \caption{Robot - Cookie + \$2}
    \label{fig:sub4}
  \end{subfigure}
  
  \caption{Pictures Used for Each Condition of the Screen Embodiment Part.}
  \label{fig:Conditions_pictures}
\end{figure*}

\paragraph{Human Stimuli.}
The two pictures (see Figures \ref{fig:sub1} and \ref{fig:sub2}) show a casually dressed man in his twenties standing. In his right hand, he holds a clear container with several cookies. His hand is slightly outstretched to show commitment and make the offer clear. The man had a \$2 coin in his left hand for the ``Cookie + \$2'' context, with the arm also positioned forward. He smiled lightly as he would have done if he was in a direct interaction. We expect this to be a better choice than a neutral face as it might have negatively influenced our results as smiling while approaching someone might respect the social norms for an interaction. The man's height is not specified, but can be approximately estimated regarding his age. The human is the same size and position in both photos.

\paragraph{Robot Stimuli.}
We took two pictures which resemble as closely as possible the interaction that will take place in physically embodied agent conditions (see \ref{experiment2_section}).
The photo shows the Nao robot (V6) on a table (see Figures \ref{fig:sub3} and \ref{fig:sub4}). In its right hand, it holds the same clear container with several cookies as used for the human photos. Its hand is slightly outstretched to show commitment and make the offer clear. For the ``Cookie + \$2'' condition, the robot had on the back of its hand a \$2 coin with the arm also positioned forward. The robot is the same size and position in both photos.

\paragraph{Recruitment and Conduction Platforms.}
Participants were recruited online via the Prolific platform. The experiment was conducted using the web-based questionnaire service Qualtrics\footnote{\url{https://www.qualtrics.com}}.

\subsubsection{Procedure}
When Prolific customers chose to participate in the study, they were redirected to the web-based questionnaire service Qualtrics. After their Prolific ID was recorded, a reCAPTCHA was displayed to exclude potential bots as their open-ended responses might fool human and AI detectors \citep{Lebrun2024}. They were then presented with an information letter outlining the risks involved, and a statement of their rights as participants (i.e., to withdraw from the study without having to justify themselves). They then had to consent to participate in the study if they wished. Otherwise, they were thanked and asked to leave the questionnaire. After giving their consent, the study was divided into four parts. The overall study lasted approximately 3 minutes.

\begin{figure*}
    \centering
    \includegraphics[width=0.7\linewidth]{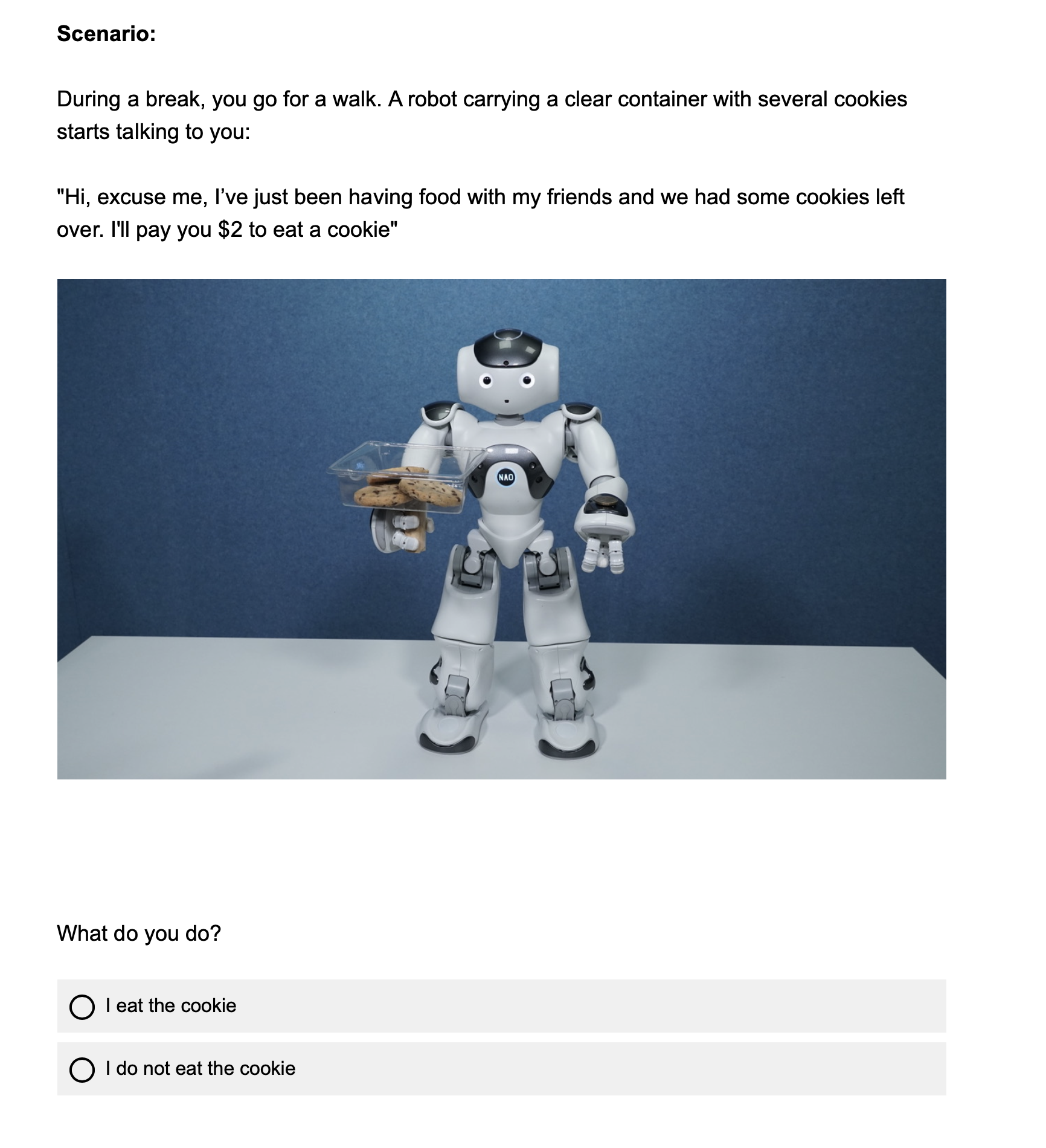}
    \caption{Example of One of the Vignettes Participants Could Meet in the Screen Embodiment Part.}
    \label{fig:vignette}
\end{figure*}

\paragraph{Vignette}
First, each participant had to read one of the four vignette where they had to imagine they were in the described situation and say what they would do. The vignettes were representing the four possible scenarios (combination of the two factors Agent and Offer) and was randomly and evenly distributed among all participants.
The vignette was presented as follows:

\begin{quote}
    During a break, you go for a walk. A [person / robot] carrying a clear container with several cookies starts talking to you: 
    
    ``Hi, excuse me, I’ve just been having food with my friends and we had some cookies left over. Would you like one? [I'll pay you 2 dollars if you eat it.]''

    What do you do?
\end{quote}
To answer the question, participants clicked on the option they thought they would have done if they were actually in the situation.
A photo representing the scenario (i.e., the person or the Nao robot with a cookie container) was shown below the text-written scenario (see \autoref{fig:vignette}).

\paragraph{Justification}
In order to better understand their choice, participants were then asked to justify their decision by writing in a text-entry box. The response participants wrote also served as an attention check for the experimenter. Participants who failed to answer this question were excluded from the data analysis.

\paragraph{Statements Ratings}
Participants rated statements on 7-point Likert scales from 1 (``Strongly disagree'') to 7 (``Strongly agree''). These were used as check Statements were
(i) ``I would eat the cookie.'' (to have more sensitive data compared to the binary question previously asked, and compare these result to the other statements),
(ii) ``The person had good intentions for offering me the cookie.'' (to check whether people attribute positive intentions to agents separated from bad intentions that are in the following statement),
(iii) ``The person had hidden reasons for offering me the cookie.'' (to check whether people imagine phantom costs), and 
(iv) ``The person gave a good explanation for why he offered the cookie.'' (to check that the HOSE model is applicable).

\paragraph{Demographics}
After having rated each statement, they were requested to provide some demographic information (i.e., gender, age, Hispanic origins, and race).

\paragraph{Debriefing}
Finally, a debriefing of the study was displayed. Participants were then automatically redirected to Prolific to confirm the completion of the study and receive their compensation on a basis of \pounds6 per hour (i.e., \pounds 0.3 for 3 minutes).

\subsection{Part B: Physical Embodiment}
\label{experiment2_section}
We conducted a between-participants design where a physically embodied agent, either a human or a robot, was making an offer that was either a cookie or a cookie associated with \$2. This part was kept as consistent as possible with the screen embodied agents part. The pre-registration for this study is available at \url{https://aspredicted.org/98X_VPG}.

\subsubsection{Sample Size Estimation}
\label{power_calculation}
The first version of the screen embodiment study was conducted to estimate the minimum required sample size for the study involving physically embodied agents. To do so, data sets were simulated with different sizes and the number of times the effect was found, at $\alpha = .05$, using the R software. The simulated data had sample sizes ranging from 20 to 80 with a step of 5. One thousand simulations were performed per sample size leading to 13,000 simulations in total. Based on these simulations, to have a power of 0.8, about n = 48 participants per condition are required to detect the same effects.  We have conducted similar analyses on the current version of the screen embodied study described above to seek the minimum required sample size. Similar analyses were conducted and suggest that n = 22 participants per condition may have been sufficient. We seek for a minimum of 48 participants per condition involving physically embodied agents to be confident in the power of our study.

\subsubsection{Participants}
251 persons at the University of Canterbury took part in the study with no prior knowledge that they would be taking part in a study.
Sufficient power is expected for this study since the sample size is in line with the power calculation of Experiment 1 (see \ref{power_calculation}).

\subsubsection{Measurements}
Participants were asked to accept or not the offer (coded as 0 if they refused and 1 if they accepted). Then participants were asked to justify their choice and provide demographic information (i.e., age, gender, and field of study).

\subsubsection{Material and Place of Study}
The study took place at the University of Canterbury. Most of the participants were students. As the photos of the screen embodied agents were shown indoors (see \autoref{fig:Conditions_pictures}), this part of the study took also place indoors in areas of the university frequented by all types of people.
The agents were not only the same used in the screen embodied agents part (see \autoref{fig:Conditions_pictures}) but also in the same position. However, here, they were physically embodied.

\paragraph{Human stimuli}
The human was a casually dressed man in his twenties. He was of average height. He was holding in his right hand a clear container with several cookies in it. For the ``Cookie + \$2'' condition, he was also holding \$2 coin in this left hand. When making his offers, he was also moving his arms forwards for people to pay attention to the offers and to show commitment. He also smiled slightly (as in \autoref{fig:sub1} and \autoref{fig:sub2})

\paragraph{Robot stimuli}
The robot used for this study was Nao\footnote{\url{http://www.aldebaran.com/fr/nao}} V6. This robot was created by Aldebaran Robotics. The robot is 58 centimetres tall with a weight of 5.4 kilograms. The Nao used for this study was blue (see \autoref{fig:Conditions_pictures}). Two robots (V5 or V6) were available as backup robots in case the main one had a problem during the experiment. Nao has 25 degrees of freedom, enabling it to perform complex movements. Several tactile sensory sensors in the hands, feet, and head, as well as microphones, speakers, cameras, and voice recognition, enable the robot to interact with people and its environment. This robot is frequently used in HRI papers because it is fully programmable using the Choregraphe\footnote{\url{http://doc.aldebaran.com/1-14/software/choregraphe/choregraphe_overview.html}} platform.
For the study, Nao was on a table for it to be at the same height level of people. It had a fixed tray in his right hand with the clear contained of cookies on it (see \autoref{fig:sub3}). If the ``Cookie + \$2'' was conducted, it also had a \$2 coin on its left hand fixed using BlueTack (see \autoref{fig:sub4}). Its left hand has a touch sensor. When the participants took the coin, the robot registered the event and thanked the participant. This was done to ensure that the experimenter will not miss one participant taking the coin as the robot's hand was not visible through the robot's camera and thus not visible to the experimenter.

\paragraph{Wizard-of-Oz Guidelines.}
Given the short time available for the recruitment but also that the resources necessary to have a robot behaving autonomously, the researchers decided to conduct the study using the Wizard-of-Oz method.
According to WoZ guidelines proposed by \citet{Riek2012}, some points need to be discussed concerning the wizard. The sole Wizard was the main researcher and therefore knew about the experiment's hypotheses. He followed a protocol to not be biased. Previous training was made in a pilot study for the wizard to be familiar with the settings. The wizard was requested to run the behavior boxes as fast as if they had performed these behaviors in real life. The Wizard script (see \autoref{appendix:b}) was very specific and consistent for all participants. In the case of the Wizard making errors, the participant's data was not recorded or a comment was reported in the data collected from the participants. However, it was expected that if some errors were going to be made, the wizard was supposed to keep interacting with the potential participants to see how they act even if they were not going to be included in the data (for not having respected the protocol).

As the robot's battery can not be left on for long without recharging, Nao was constantly being charged via its charger. The cabling is not expected influence the perceived autonomy of the robot \citep{Roesler2023} Both the computer and the robot were connected to the same WiFi using a router. While the computer was physically connected to the router using an Ethernet cable, Nao was connected in wireless. This configuration was chosen to make the robot's non-autonomy less obvious and allow the Wizard operating several meters from  the robot. In this way, the experimenter controlled the robot in real-time via Choregraphe. Also, to not create a habituation effect or people guessing a study was running since people would have seen the robot continuously at the same place in the same building, the robot changed its location regularly. To make the interaction more natural and that people perceive the robot answered and behaved adequately with participants, the experimenter had access in real-time to the robot's cameras on the Choregraphe platform. The Wizard did not need to look at people in real-life but used the robot's cameras for it. When an action was required (e.g., the robot had to extend its arm and talk), the experimenter just had to double-click on the corresponding box ``start'' button.

\subsubsection{Procedure}
The study design followed the one of \citet{vonasch2022} as much as possible, except that whereas in the original study the agent (a person) approached participants, this was not done in the present study because the robot could not walk. Instead, the robot called out to potential participants as they walked by. To keep the procedure as similar as possible between conditions, the human in this study did the same. There was one other difference between this study and the original. In \citep{vonasch2022}, only individuals were contacted. In the current study, groups of people were also contacted because because alone people were usually distracted by their phones or listening to music.
Participants had no prior knowledge that they would be taking part in a study. Thus, their informed consent was not obtained prior to their participation. The experiment lasted approximately 2 minutes.

\paragraph{Experimenter Position}
The experimenter was positioned several meters from the robot, but close enough to hear any conversations people might have with the robot. He sat at a table with his back to the participants who would be interacting with the robot. He used the computer with Choregraphe. So, not only did he look like a student studying his lessons, but he also had a good view of the situation, without the robot being in the way.

\paragraph{Scenario (Before Consent)}
Participants were called out by the human or the robot positioned on a table. The conditions shown in Figures \ref{fig:Conditions_pictures} represent the situation encountered by the participants. The robot is positioned on the table, holding a container containing several cookies. We chose to position the robot on a table so that it would be at approximately head height with the participants. The robot had to be approached by the participants as it was unable to walk correctly  around the campus. To maintain consistency between conditions, the human also called out to the participants without moving towards them as well. In both conditions, human and robot, the agent will say the following sentence:

\begin{quote}
    ``Hi, excuse me, I’ve just been having food with my friends and we had some cookies left over. Would you like one? [I'll pay you 2 dollars if you eat it.]''
\end{quote}

\noindent In the ``Cookie'' condition, participants only had to take the cookie or not. In the ``Cookie + \$2'' condition, participants had to accept both offers, the cookie and a \$2 coin. In the robot cookie + \$2 condition, when the participant took the coin, the robot automatically thanked him. Except this behavior, every robot behavior were triggered using the Wizard of Oz protocol.

It's important to note that the ``robot'' condition was setup in accordance with the Wizard of Oz protocol. The robot displayed full autonomy while the experimenter controlled the robot remotely. The experimenter kept an eye on the situation. In this way, the experimenter had access to the live stream on his computer, thanks to Nao's cameras positioned at the forehead and mouth. The wizard could hear the conversations participants had with the robot. To do so, the Wizard was hidden from people's view when they were interacting with the robot but close enough to hear what they were saying to the robot. The video was not recorded, this simply enabled the experimenter to control the robot's behavior in the best possible way, to achieve a more natural interaction, and to know what the robot should do in real time.

\paragraph{Justification and Demographics (After Consent)}
After participants accepted or rejected the offer, the experimenter (in the human conditions) or students who volunteered to help (in the robot conditions) explained them this was a study in psychology. They received a debriefing of the study and had the option of withdrawing their participation without affecting the compensation received. When participants agreed to keep their participation available for the study, the experimenter asked them to explain their choice and to provide some demographic information (i.e., gender, age, field of study).

\section{Results}
\subsection{Descriptive Analyses}
\subsubsection{Part A: Screen Embodiment}
Among the 481 participants, 15 participants have been removed from the data analyses because did not meet the attention check justification (n = 1) or had dietary restrictions (n = 14) leading to a final sample size of 466 participants. Their age ranged from 18 to 78 $(M = 38.31, SD = 13.01, Median = 36)$.
The gender distributions is as follows: 232 were males, 228 were females, and 6 said having a different gender. Genders were almost equally distributed between conditions (see \autoref{tab:gender_condition}). All participants were American. 9\% (n = 42) declared having a Spanish, Latin, or Hispanic origin. Race is distributed as follows: 73.39\% of participants were White or Caucasian, 11.80\% were Black or African American, 1.72\% were American Indian/Native American or Alaska Native, 14.81\% were Asian, 0.43\% were Native Hawaiian or Other Pacific Islander, 2.36\% were had another race and 0.86\% of the participants preferred to not answer the question.

\begin{table}[htbp]
\centering
\caption{Gender Counts per Screen Embodied Agent and Offer}
\label{tab:gender_condition}
\begin{tabular}{lcccc}
\toprule
  & \multicolumn{2}{c}{Human} & \multicolumn{2}{c}{Robot} \\ \cmidrule{2-5}
  & Cookie   & CD   & Cookie   & CD   \\ \cmidrule{2-5}
Male   & 53       & 62             & 57       & 60             \\
Female & 61       & 56             & 54       & 57             \\
Other  & 3        & 0              & 2        & 1              \\ \bottomrule
\end{tabular}
\footnotetext{Note: CD: Cookie + \$2.}
\end{table}

\subsubsection{Part B: Physical Embodiment}
27 participants were removed from the data analyses according to the following criteria: 11 participants did not accept the cookie due to dietary restrictions, 9 guessed it was a study, 5 did not follow the protocol or misunderstood the interaction or the offer, one person ran away when they saw the experimenter approaching them and therefore did not consent, and one participant was a university student but still minor. The final sample was composed of 224 participants.

The participants' age in the final sample ranged from 18 to 61 $(M = 22.5, SD = 6.09, Median = 20)$. While 2.68\% (n = 6) were University Staff, 97.32\% (n = 218) were students. The gender distribution is shown in \autoref{tab:gender_condition_inperson}. Their field of study was distributed as follows: 62 in Engineering, 59 in Science, 35 in Commerce and Related, 17 in Law, 16 in Teaching and Language, 4 in Technology, and 25 had another field of study.

\begin{table}[htbp]
\centering
\caption{Gender Counts per Physically Embodied Agent and Offer}
\label{tab:gender_condition_inperson}
\begin{tabular}{lcccc}
\toprule
  & \multicolumn{2}{c}{Human} & \multicolumn{2}{c}{Robot} \\ \cmidrule{2-5}
  & Cookie   & CD   & Cookie   & CD   \\ \cmidrule{2-5}
Male   & 38       & 32             & 26       & 30             \\
Female & 16       & 22             & 29       & 30             \\
Other  & 0        & 0              & 1        & 0              \\ \bottomrule
\end{tabular}
\footnotetext{Note: CD: Cookie + \$2.}
\end{table}

\subsection{Predictors of Offer Acceptance}
\begin{figure*}
\begin{center}
\includegraphics[width=\linewidth]{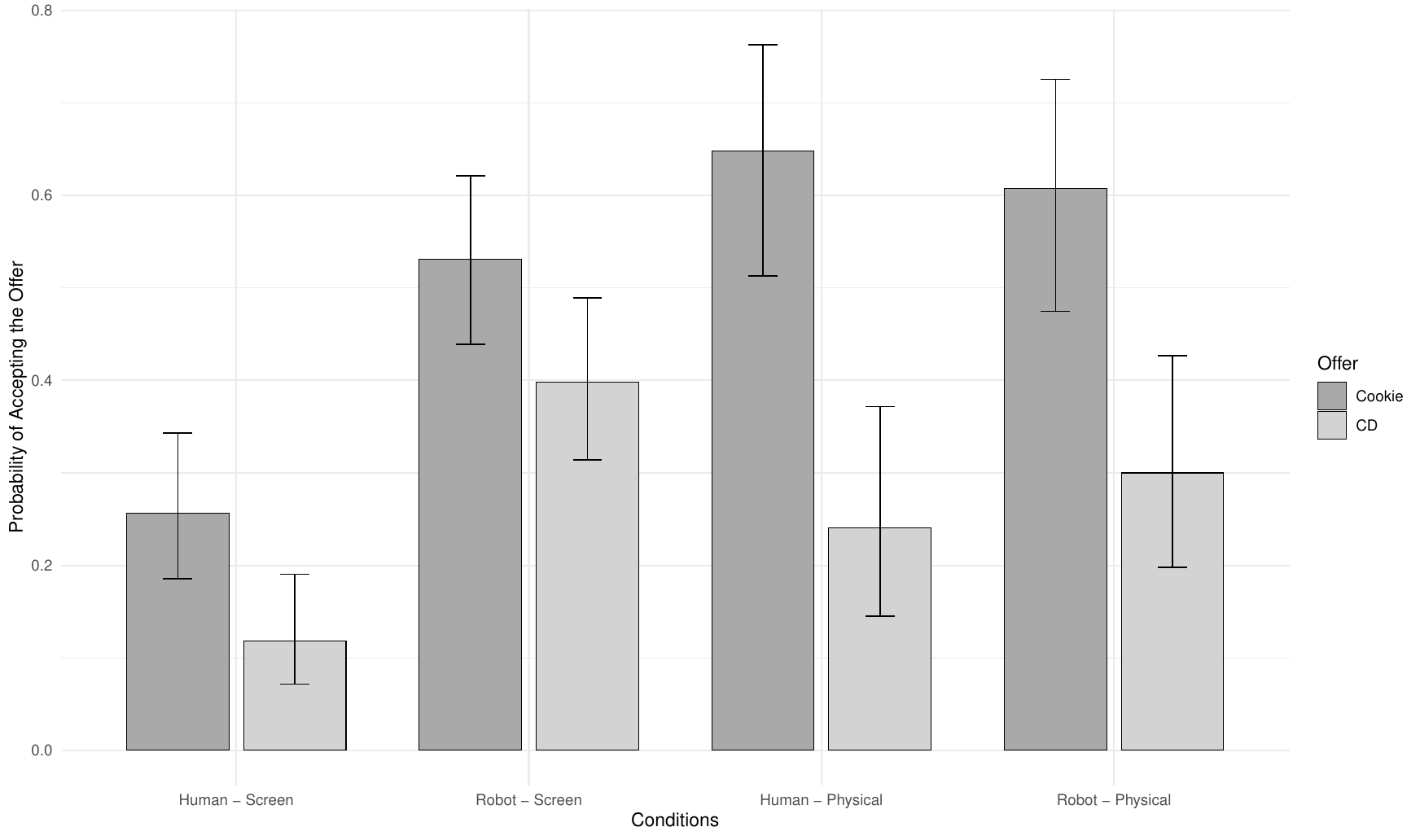}
\end{center}
\footnotesize{Note: CD corresponds to the Cookie + \$2 condition. Errors bars are 95\% confidence intervals (95\%CI).}
\caption{Plot of the Probability of Accepting the Offer as a Function of the Agent, the Offer, and the Agent Embodiment.}
\label{fig:OverallModel}
\end{figure*}

Three-way interactions logistic regression analyses were conducted using the  R software to examine the relationship between the Agent, their Embodiment, and the Offer the Agents made, on the probability of Rejecting (0) or Accepting (1) the Offer. The Agent was either a Human (0) or a Robot (1). They were either screen (0) or physically (1) embodied. The Offer was either a Cookie (0) or a Cookie + \$2 (1). When agents were physically embodied , 12 participants took the cookie but not the coin in the ``Cookie + \$2'' offer. They were not removed from analyses and were counted as 1 (i.e., ``Yes, accepted the offer'') since they chose to eat the cookie, which was the purpose of the analyses and research questions. Note that excluding these participants from data analyses provided the same results overall.
For clarity, ``accepting or rejecting the offer'' will still be used to explain the results. The model equation is shown in in \autoref{eq:overall_model}). The output of the model is shown in in \autoref{tab:CoefficientsOverallModel} and the plot of the data per condition in \autoref{fig:OverallModel}. The logistic regression model shows significant results for each factor. 95\% confidence intervals (95\%CI) are shown in the figure. Any overlap of these CI between any conditions indicated the absence of significant differences between these conditions.

\begin{figure*}
\begin{equation}
\begin{aligned}
\text{logit}(\mathbb{P}(\text{Choice} = 1)) = & \beta_0 + \beta_1 \cdot \text{Agent} + \beta_2 \cdot \text{Offer} + \beta_3 \cdot \text{Embodiment} \notag \\
&\quad + \beta_4 \cdot (\text{Agent} \cdot \text{Offer}) + \beta_5 \cdot (\text{Agent} \cdot \text{Embodiment}) + \beta_6 \cdot (\text{Offer} \cdot \text{Embodiment}) \notag \\
&\quad + \beta_7 \cdot (\text{Agent} \cdot \text{Offer} \cdot \text{Embodiment})
\label{eq:overall_model}
\end{aligned}
\end{equation}    
\end{figure*}

\begin{table*}[ht]
\caption{Coefficients of the Three-Way Interactions Logistic Regression Model.}
\begin{center}
        \begin{tabular}{lcccr}
        \toprule
          Effect & Estimate & Standard Error & z value & $Pr(>|z|)$ \\ \midrule
         (Intercept) & -1.0647 & 0.2117 & -5.029 & $< .001^*$ \\
         Agent & 1.1888 & 0.2835 & 4.193 & $< .001^*$ \\
         Offer & -0.9406 & 0.3548 & -2.651 & $< .001^*$ \\
         Embodiment & 1.6756 & 0.3550 & 4.720 & $< .001^*$ \\
         Agent $\times$ Offer & 0.4040 & 0.4436 & 0.911 & .362\phantom{*} \\
         Agent $\times$ Embodiment & -1.3644 & 0.4862 & -2.806 & .005$^*$\\
         Offer $\times$ Embodiment & -0.8189 & 0.5553 & -1.475 & .140\phantom{*} \\
         Agent $\times$ Offer $\times$ Embodiment & 0.0729 & 0.7304 & 0.100 & .921\phantom{*} \\
         \bottomrule
    \end{tabular}
    \end{center}
\footnotesize{Note: *: $p < .05$}
\label{tab:CoefficientsOverallModel}
\end{table*}

Main effects are directly obtained from \autoref{tab:CoefficientsOverallModel}. The two-way interaction Agent $\times$ Embodiment was also significant ($p = .005$). To understand all the effects better, post-hoc analyses were conducted using the \textit{emmeans} package\footnote{\url{https://cran.r-project.org/web/packages/emmeans/index.html}}. Tests were performed on the log odds ratio scale.

\subsubsection{Offer: Cookie and Cookie + \$2}
The odds of accepting the offer were 2.56 times higher ($CI 95\% [1.30, 5.27]$) when the offer was a cookie compared to when it was a cookie associated with \$2.
For each Agent $\times$ Embodiment pairwise comparisons, the odds of accepting the Offer were higher when the offer was a cookie compared to a cookie associated with \$2. The odds ratio (OR = Cookie/(Cookie + \$2)) of each pairwise comparison were the followings. When the agent was a human screen embodied, the OR of the offer was 2.56 ($p = .008$) whereas the OR of the offer for a physically embodied human was 5.81 ($p < .001$). When the agent was a screen embodied agent, the OR was 1.71 ($p = .044$) whereas the OR for the physically embodied robot was 3.61 ($p = .001$).

\subsubsection{Agent: Human and Robot}
The odds of accepting the offer were 3.28 times higher ($CI 95\% [1.89, 5.58]$) when the Robot was making the offer compared to the Human making an offer.
While no significant difference has been found between the Human and the Robot when they were physically embodied, neither for the Cookie condition ($p = .657$) nor the Cookie + \$2 condition ($p = .478$), the odds of accepting the offer were higher for the screen embodied robot compared to the screen embodied human when the offer was a cookie (OR = 3.27, $p < .001$) but also a cookie + \$2 (OR = 4.93, $p < .001$).

\subsubsection{Embodiment: Screen and Physical}
The odds of accepting the offer were 5.34 times higher ($CI 95\% [2.70, 10.90]$) when agents were physically embodied compared to screen embodied.
Only the embodiment of the human agent influenced the odds of accepting the offer. The odds of accepting the cookie were 5.35 times higher when the human was physically embodied compared to screen embodied ($p < .001$). The odds of accepting the cookie associated with a monetary outcome were 2.35 times higher for the physically embodied human compared to the screen embodied one. These differences have not been found when the agent was a human ($p = .349$ for the cookie condition and $p = .199$ for the cookie + \$2 condition).

\subsubsection{Gender Analyses}
Exploratory analyses have been conducted to explore potential gender differences for both embodiment levels on the Choice dependent variable.
When agents were \textit{screen} embodied, an ANOVA was conducted ($M_{male} = 0.39, SD_{male} = 0.03, M_{female} = 0.25, SD_{female} = 0.03, M_{other} = 0.67, SD_{other} = 0.19$) and showed a Gender effect, $F(2,463) = 6.77, p = .001, \eta^2 = 0.028$. Post-hoc analyses showed that males were more likely to accept the offer than females, $t(463) = 3.20, p = .004, d = 0.30, 95\%CI[0.11, 0.48]$.

When agents were \textit{physically} embodied, analyses were conducted using a chi-square test to explore potential Choice differences between males ($M = 0.50, SD = 0.50$) and females ($M = 0.37, SD = 0.49$). The third gender (identified as ``Other'') was removed from data analyses as only one participant declared having a non-binary gender. Data analyses did not show any significant difference between these two groups, ($\chi^2(1) = 3.69, p = .055$) although notably if there was a gender difference, it was in the same direction as in the screen embodied condition.

\subsection{Decision-Making Criteria}
Two coders individually categorized the participants' justifications to understand the criteria participants were using to make their decisions. The inter-reliability between both coders was satisfactory ($\kappa = 0.78$ and $\kappa = 0.71$ respectively for the screen and physically embodied agents studies). Disagreement between both coders was resolved by discussing until a complete agreement ($\kappa = 1.00$) was achieved. The criteria and their sum are shown in \autoref{tab:criteria}.

Participants mainly justified their choice using 10 criteria gathered into seven categories. These categories and their definitions (used for the coders to categorize the data) were: (i) Phantom Costs (i.e., suspicion, something is weird, the cookie might be poisoned etc), (ii) Excitement (i.e., people being happy because of the cookie or the dollars directly such as getting free food), (iii) Hunger (i.e., being or not hungry), (iv) Agent (i.e., everything related to the Agents such as trusting them, considering them as cute or dodgy, a kind gesture etc), (v) Sanitation (i.e., not clean, cookies not sealed etc), (vi) Food from Strangers (i.e., not accepting food from stranger as a rule of thumb), and (vii) Etiquette (i.e., altruism, participants who felt bad accepting the money, accepting the offer to be polite, etc). Some other information were provided by the participants and are categorized as ``Other.''

\begin{table*}[]
\begin{center}
\caption{Sum of the Decision Criteria Specified by the Participants to Make Their Decisions According to the Agent, the Embodiment, and the Offer}
\label{tab:criteria}
\begin{tabular}{lrrrrrrrr}
\toprule
                    & \multicolumn{4}{c}{Screen}                 & \multicolumn{4}{c}{Physical}               \\
\cmidrule(lr){2-5} \cmidrule(lr){6-9}
                    & \multicolumn{2}{c}{Human} & \multicolumn{2}{c}{Robot} & \multicolumn{2}{c}{Human} & \multicolumn{2}{c}{Robot} \\ \cmidrule(lr){2-5} \cmidrule(lr){6-9}
Criteria            & Cookie   & CD   & Cookie   & CD   & Cookie   & CD   & Cookie   & CD   \\
\midrule
Phantom Costs       & 35         & 82               & 22         & 51               &  4        &  22              &  7        &   28             \\
Excitement          & 19         & 13               & 42         & 41               & 19         & 7               &  13        & 11               \\
Hungry              & 0         & 1               & 3         & 0               & 6         &  3              &  5        &  2              \\
Not Hungry          & 0         & 0               & 0         & 1               & 8         &  6              &  6        &     5           \\
Good Agent          & 16         &  0              & 17         & 5               &  18        &  0              &  19        &   6             \\
Bad Agent           &  5        &  4              &  7        &   17             &  2        &    7            &  4        &  8              \\
Sanitation          &  12        & 5               & 13         &   14             &  2        &  0              & 4         &   8             \\
Food from Strangers &  69        & 57               &  24        &  14              &   2       & 3               & 1         &   0             \\
Etiquette           & 11         &  0              &  4        &  0              &  1        &  5              &  2        &   4             \\
Other               & 24         &  5              & 23         &  16              &   6       & 16               &   9       & 13               \\
\bottomrule
\end{tabular} \\
\end{center}
\footnotesize{Note: CD corresponds to the Cookie + \$2 condition. Values are the number of participants who reported such criteria in their choice justification.}
\end{table*}

\begin{figure*}
  \centering
  \begin{subfigure}[b]{0.3\textwidth}
    \includegraphics[width=\textwidth]{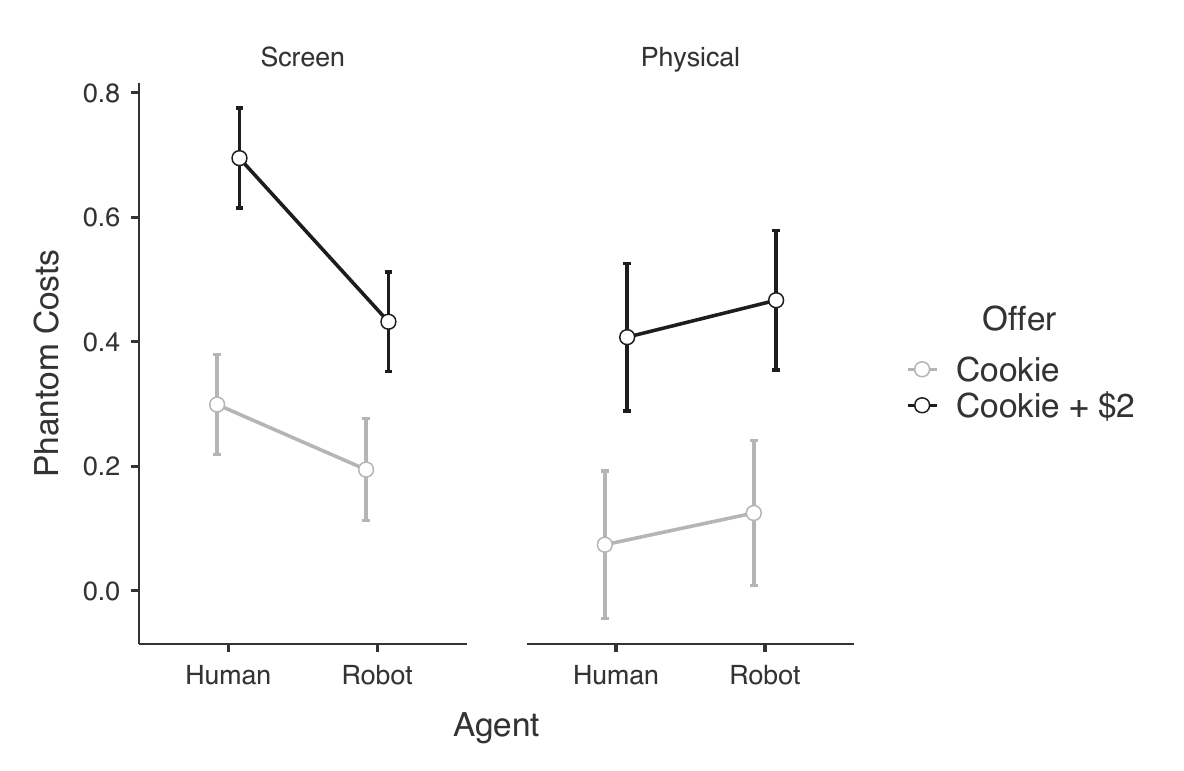}
    \label{subfig:phantom_cost}
  \end{subfigure}
  \hfill
  \begin{subfigure}[b]{0.3\textwidth}
    \includegraphics[width=\textwidth]{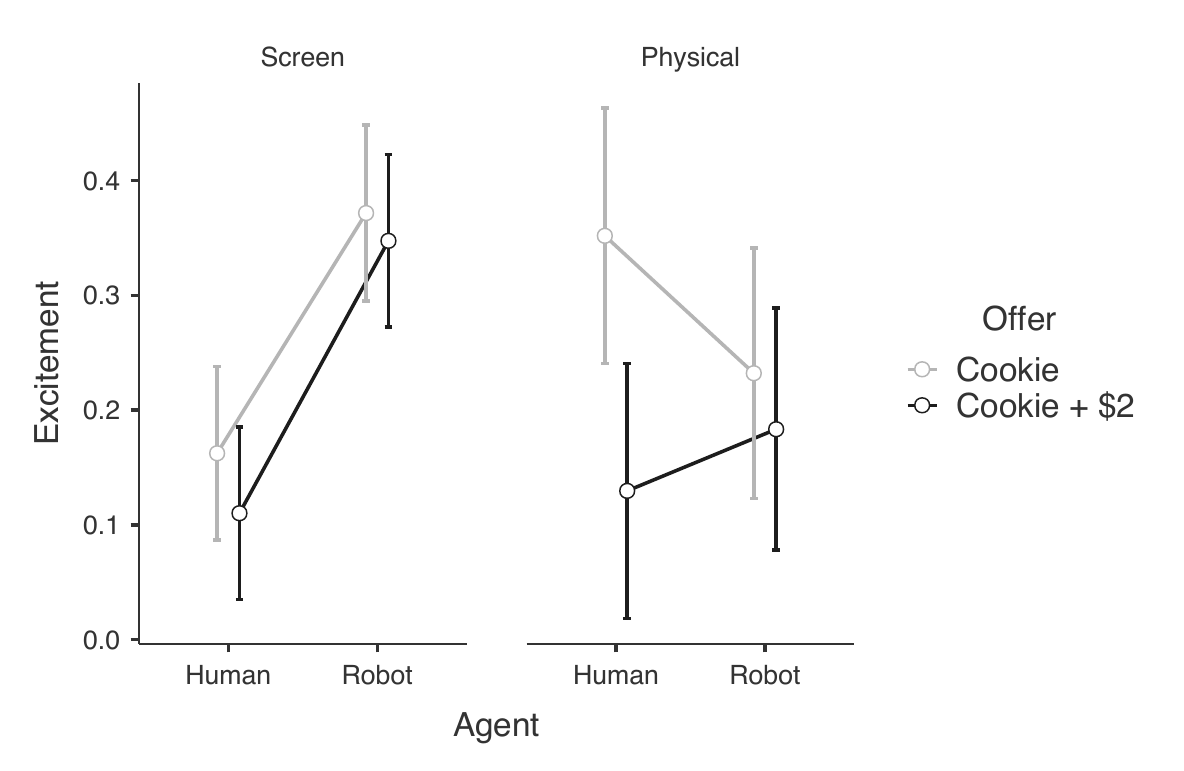}
    \label{fig:excitement}
  \end{subfigure}
  \hfill
  \begin{subfigure}[b]{0.3\textwidth}
    \includegraphics[width=\textwidth]{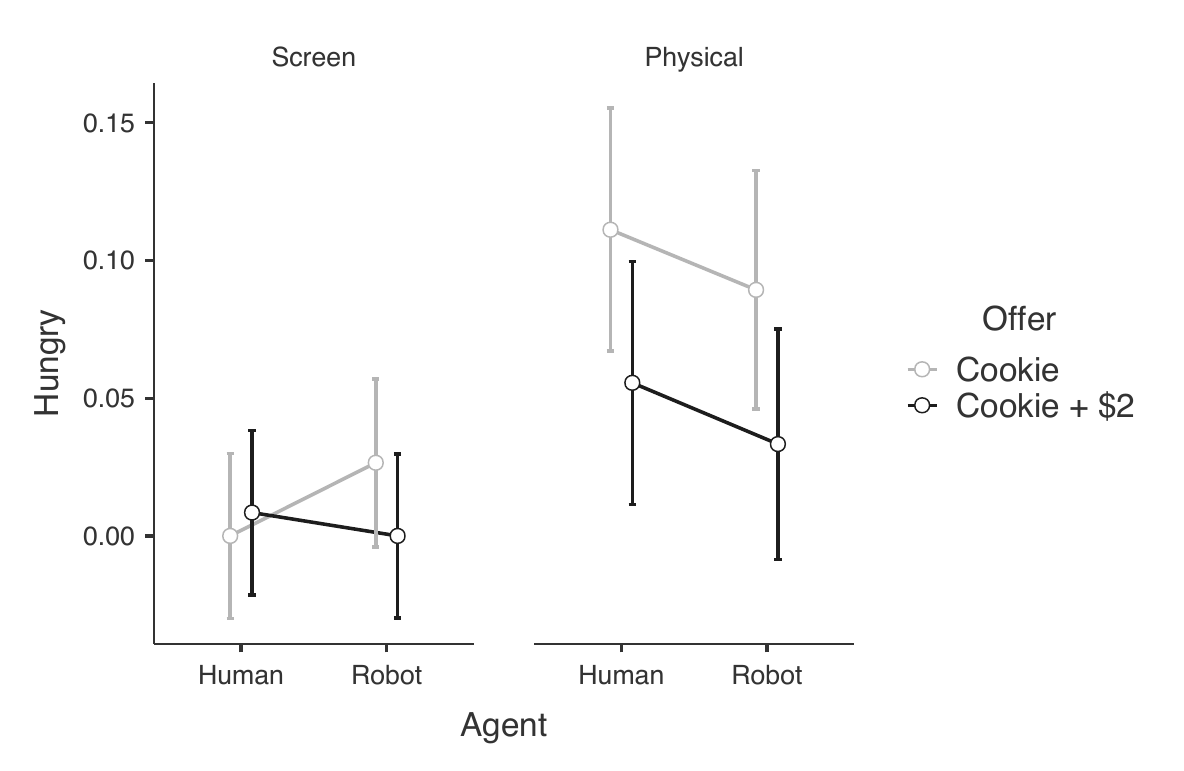}
    \label{fig:hungry}
  \end{subfigure}
  \hfill
  \begin{subfigure}[b]{0.3\textwidth}
    \includegraphics[width=\textwidth]{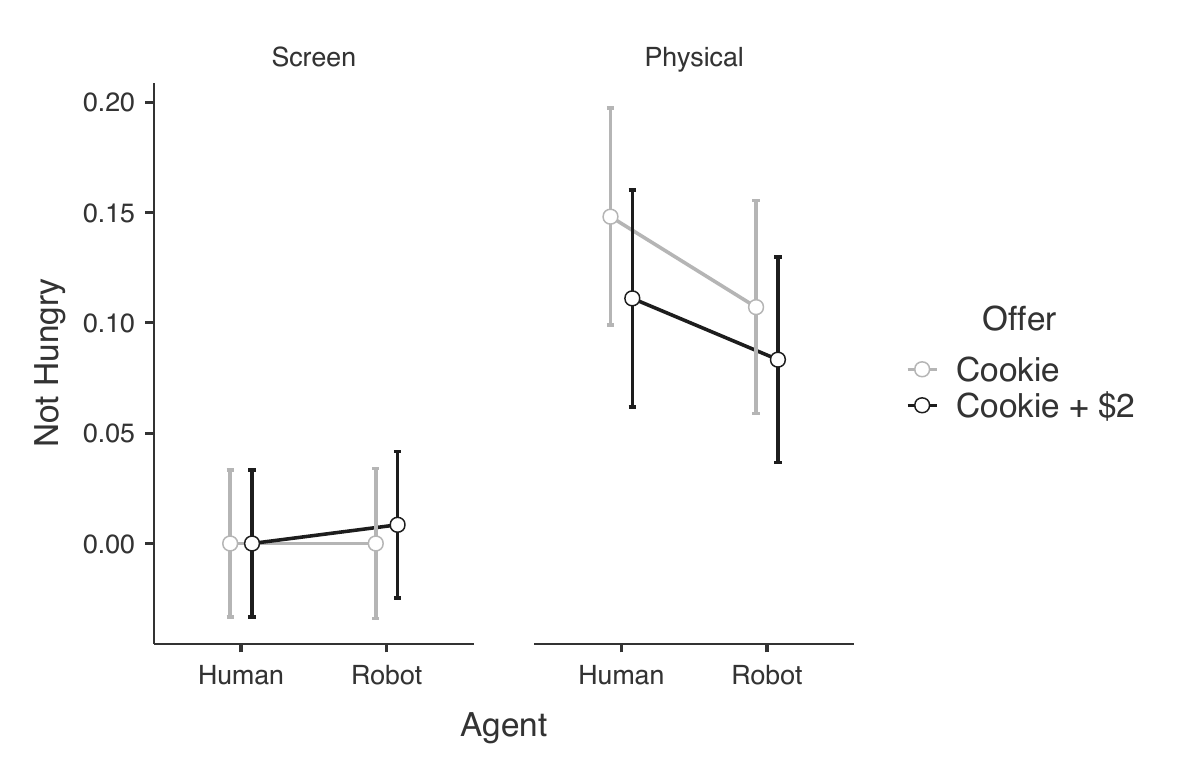}
    \label{fig:not_hungry}
  \end{subfigure}
  \hfill
    \begin{subfigure}[b]{0.3\textwidth}
    \includegraphics[width=\textwidth]{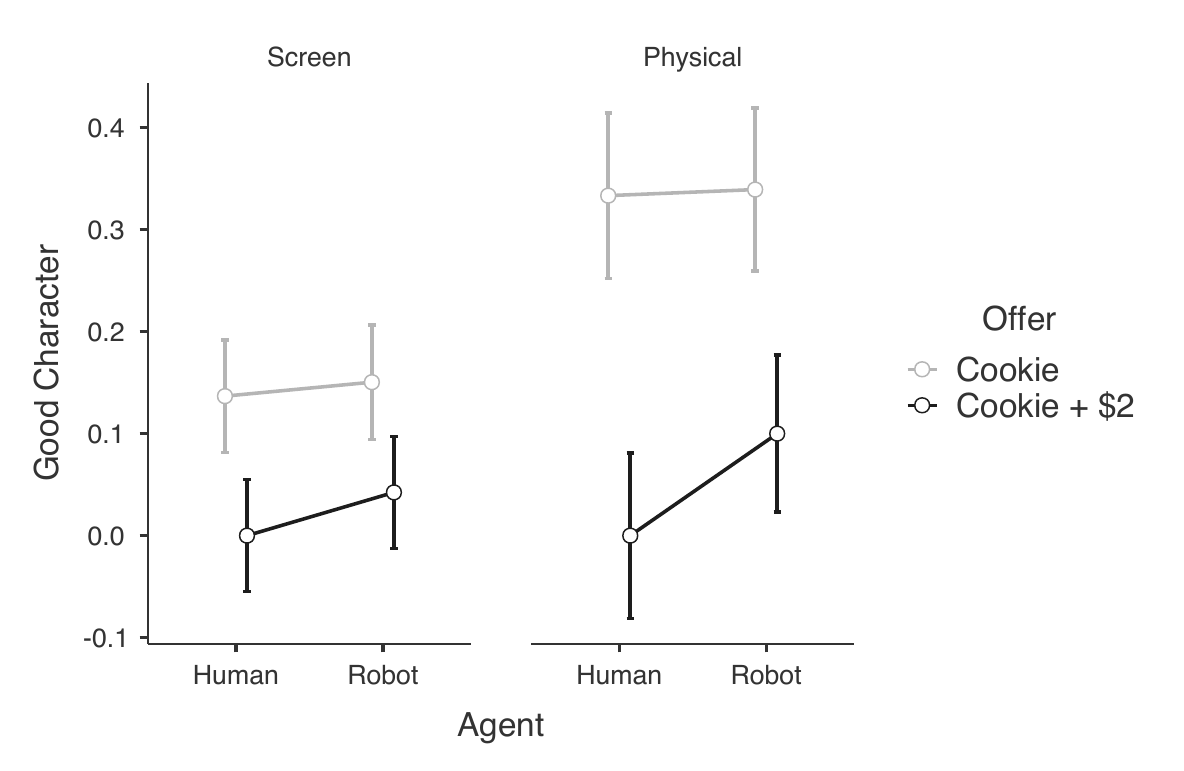}
    \label{fig:good_character.pdf}
  \end{subfigure}
  \hfill
  \begin{subfigure}[b]{0.34\textwidth}
    \includegraphics[width=\textwidth]{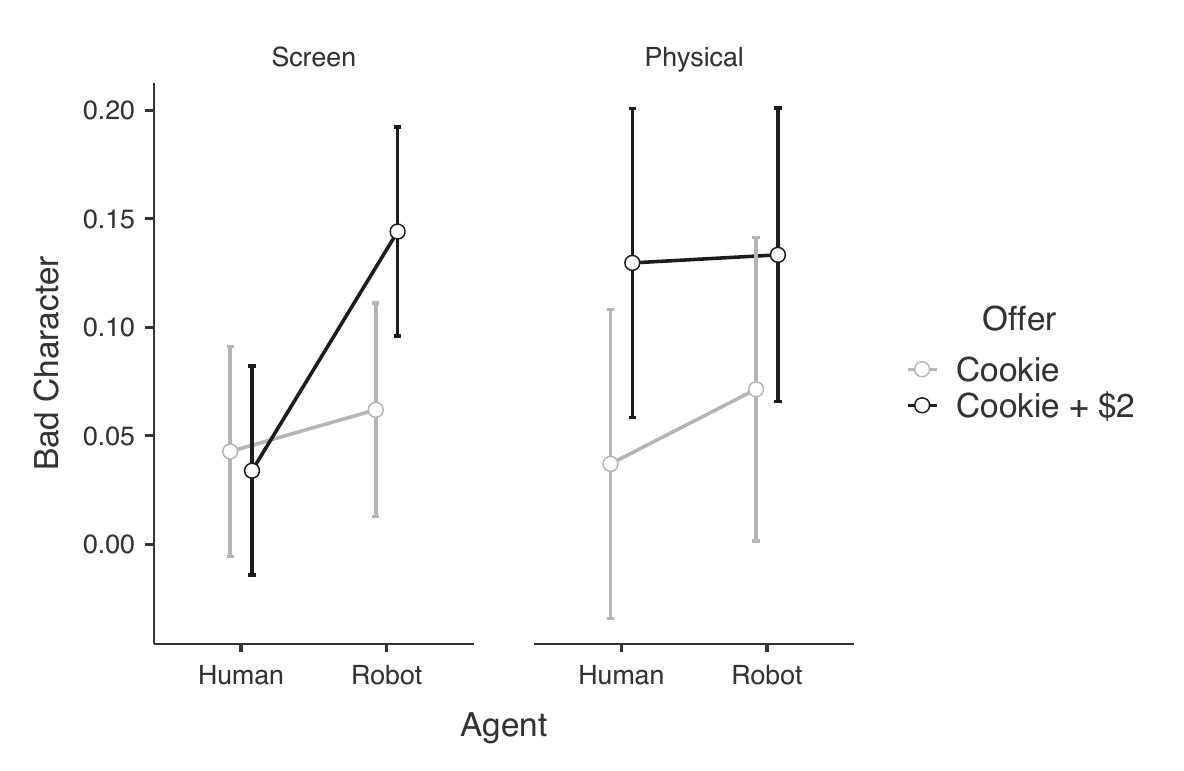}
    \label{fig:bad_character}
  \end{subfigure}
  \hfill
    \begin{subfigure}[b]{0.3\textwidth}
    \includegraphics[width=\textwidth]{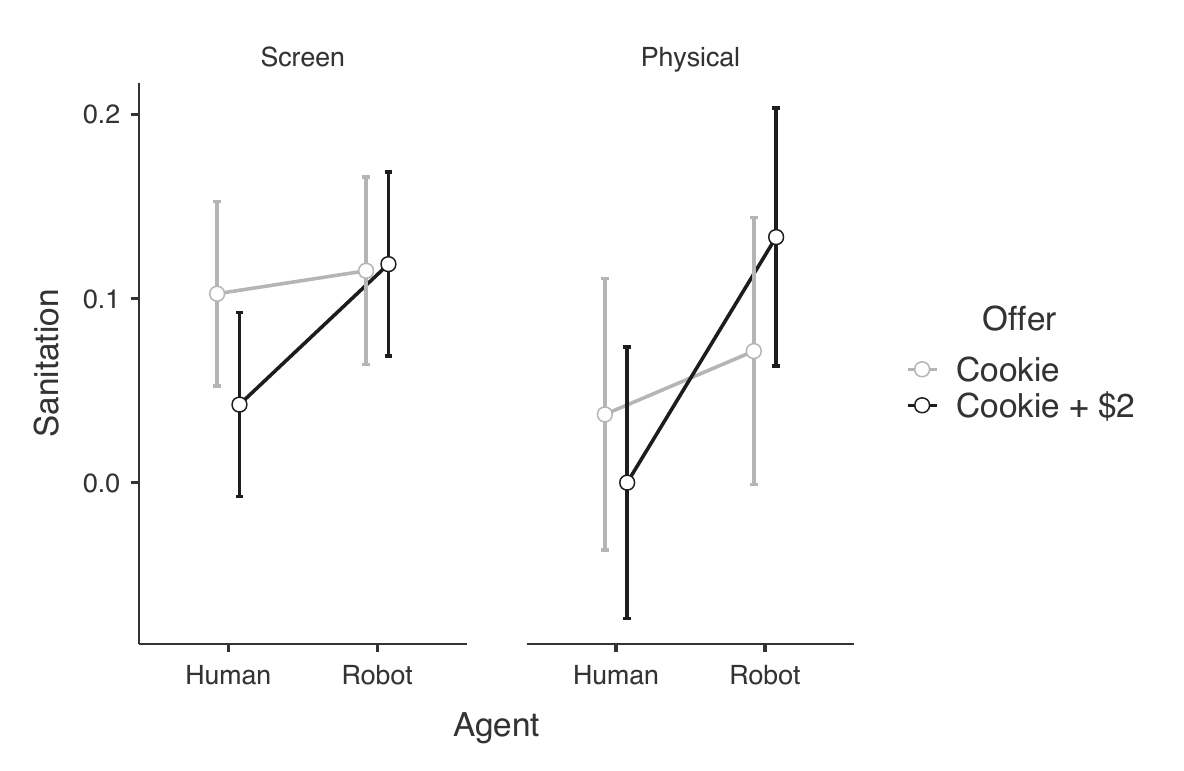}
    \label{fig:sanitation}
  \end{subfigure}
  \hfill
  \begin{subfigure}[b]{0.3\textwidth}
    \includegraphics[width=\textwidth]{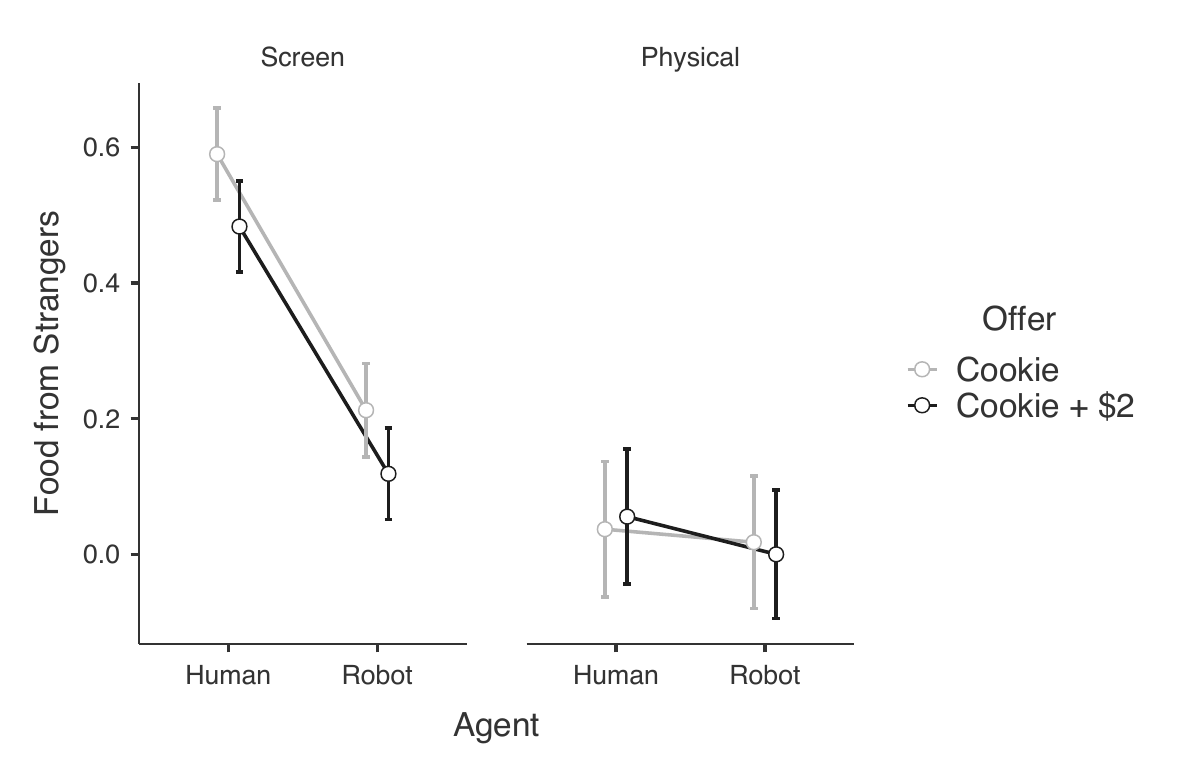}
    \label{fig:stranger}
  \end{subfigure}
  \hfill
    \begin{subfigure}[b]{0.3\textwidth}
    \includegraphics[width=\textwidth]{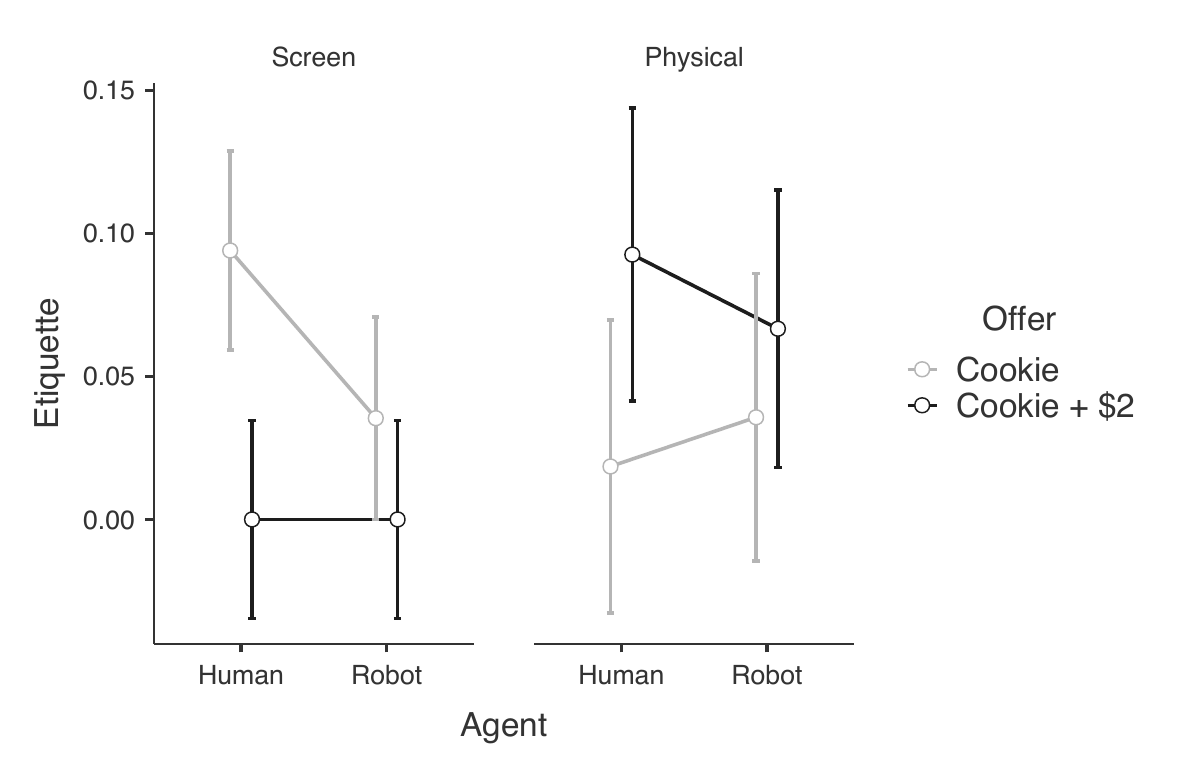}
    \label{fig:etiquette}
  \end{subfigure}
  \hfill

  \caption{Graphs Representing Each Criteria as a Function of the Agent, the Offer, and the Embodiment.}
  \label{fig:criteria}
\end{figure*}

Analyses of Variance (ANOVA) have been conducted to explore the effect of the Agent, the Offer, and the Embodiment on the perception of each of the ten criteria described in \autoref{tab:criteria}. Plots are shown in \autoref{fig:criteria}. Results are discussed below. Student t-tests with Tukey correction have been conducted to understand better the direction of the effects and the interaction effects.

\begin{itemize}
    \item[\scalebox{0.8}{$\bullet$}] \textbf{Phantom Costs}. 
    There is a main effect of the Offer ($F(1,682) = 82.24, p < .001, \eta^2 = 0.103$) and the Embodiment ($F(1,682) = 14.42, p < .001, \eta^2 = 0.018$), as well as an interaction effect Agent $\times$ Embodiment ($F(1.682) = 10.95 , p < .001, \eta^2 = 0.014$) on the report of phantom costs by participants.
    More specifically, participants reported more often phantom costs when (i) the cookie was associated with a monetary outcome ($mean = 0.52, SD = 0.50$) compared to a cookie alone ($mean = 0.20, SD = 0.40$), $t(682) = -9.07, p < .001, d = 0.74$, but also when (ii) agents were screen embodied ($mean = 0.41, SD = 0.49$) compared to physically embodied ($mean = 0.27, SD = 0.45$), $t(682) = 3.80, p < .001, d = 0.31$.
    Post-hoc analyses have also been conducted to understand better the Agent $\times$ Embodiment interaction effect. Participants reported more  phantom costs when the human was screen embodied compared to physically embodied ($t(682) = 4.97, p < .001, d = 0.58$) but also when the screen embodied human was making the offer compared to a screen embodied robot ($t(682) = 4.47, p < .001, d = 0.41$) or a physically embodied robot ($t(682) = 4.00, p < .001, d = 0.45$).
    
    \item[\scalebox{0.8}{$\bullet$}] \textbf{Excitement}. 
    The results showed main effects for the Agent ($F(1, 682) = 7.90, p = 0.005, \eta = 0.011$) and Offer ($F(1, 682) = 6.59, p = 0.010, \eta = 0.009$) factors as well as a Agent $\times$ Embodiment interaction effect ($F(1,682) = 14.34, p < .001, \eta^2 = 0.020$) on the Excitement reported by the participants about the offer.
    More specifically, people reported more excitement about the offer when (i) it was offered by a robot ($mean = 0.31, SD = 0.46$) compared to a human ($mean = 0.17, SD = 0.38$), $t(682) = -2.81, p = .005, d = 0.23$, but also when (ii) the offer was a cookie alone ($mean = 0.27, SD = 0.45$) compared to a cookie + \$2 ($mean = 0.21, SD = 0.41$), $t(682) = 2.57, p = .010, d = 0.21$.
    Post-hoc analyses have also been conducted to understand better the Agent $\times$ Embodiment interaction effect. Participants reported more often being excited about the offer when the screen embodied robot was making the offer ($mean = 0.36, SD = 0.03$) compared to when it was a screen embodied human ($mean = 0.14, SD = 0.03$), $t(682) = -5.79, p < .001, d = 0.54$. People also reported more excitement about the offer when the robot was screen embodied ($mean = 0.36, SD = 0.03$) compared to physically embodied ($mean = 0.21, SD = 0.04$), $t(682) = 3.21, p = .008, d = 0.37$.

    \item[\scalebox{0.8}{$\bullet$}] \textbf{Hunger}. 
    Concerning the fact of being hungry, the results showed a main effect of the Offer ($F(1,682) = 5.82, p = .016, \eta^2 = 0.008$) and the Embodiment ($F(1,682) = 22.42, p < .001, \eta^2 = 0.031$) on the perception of hunger by participants.
    More specifically, people reported being hungry more often when the offer was a cookie alone ($mean = 0.04, SD = 0.2$) compared to when it was associated with money ($mean = 0.02, SD = 0.13$), $t(682) = 2.41, p = .016, d = 0.20$. People also reported more often being hungry when agents were physically embodied ($mean = 0.07, SD = 0.26$) compared to screen embodied ($mean = 0.009, SD = 0.09$), $t(682) = -4.74, p < .001, d = 0.39$.
    Concerning the fact of not being hungry, the results showed a main effect of the Embodiment ($F(1,682) = 54.33, p < .001, \eta^2 = 0.073$). More specifically, participants reported more often not being hungry when agents were physically embodied ($mean = 0.11, SD = 0.32$) compared to screen embodied ($mean = 0.002, SD = 0.05$), $t(682) = -7.37, p < .001, d = 0.6$.

    \item[\scalebox{0.8}{$\bullet$}] \textbf{Character}. 
    Results showed a significant effect of the Offer ($F(1,682) = 68.31, p < .001, \eta^2 = 0.087$) and the Embodiment ($F(1,682) = 20.07, p < .001, \eta^2 = 0.026$) as well as their interaction ($F(1,682) = 10.98, p < .001, \eta^2 = 0.014$) on positive perceptions of the agent.
    More specifically, people reported more positive information concerning the agent when the offer was a cookie alone ($mean = 0.21, SD = 0.41$) compared to a cookie associated with \$2 ($mean = 0.03, SD = 0.18$), $t(682) = 8.26, p < .001, d = 0.67$. More positive information about the agent were reported by participants when the agent was physically embodied ($mean = 0.19, SD = 0.40$) compared to screen embodied ($mean = 0.08, SD = 0.27$), $t(682) = -4.48, p < .001, d = 0.36$.
    Post-hoc analyses have also been conducted to understand better the Offer $\times$ Embodiment interaction effect. All effects were significant except the one involving the embodiment effect with the Cookie + \$2 offer ($p = .840$).
    More specifically, participants reported more often positive information about the physically embodied agent offering a cookie ($mean = 0.34, SD = 0.03$) compared to screen embodied agents making the same offer ($mean = 0.14, SD = 0.02$), $t(682) = -5.47, p < .001, 0.63$. Screen embodied agents offering a cookie alone were also received more positive information from participants ($mean = 0.14, SD = 0.02$) compared to when they offered a cookie associated with \$2 ($mean = 0.02, SD = 0.02$), $t(682) = 4.34, p < .001, d = 0.40$. Participants also reported more often positive perceptions of the screen embodied agents offering a cookie alone ($mean = 0.14, SD = 0.02$) compared to the physically embodied agents offering a cookie and \$2 ($mean = 0.05, SD = 0.03$), $t(682) = 2.69, p = .037, d = 0.31$. The opposite conditions are also different: participants reported more often positive perceptions of the physically embodied agents offering a cookie alone ($mean = 0.34, SD = 0.03$) compared to the screen embodied agents offering a cookie and \$2 ($mean = 0.02, SD = 0.02$), $t(682) = 8.98, p < .001, d = 1.04$.
    Results showed an effect of the Offer factors on the negative perceptions about the agent that participants reported, $F(1,682) = 6.90, p = .009, \eta^2 = 0.010$. The Agent factor is barely not significant, $F(1,682) = 3.73, p = .054$.
    More specifically, participants reported more often negative information about the agent when the offer was a cookie + \$2 ($mean = 0.10, SD = 0.30$) compared to a simple cookie ($mean = 0.05, SD = 0.22$), $t(682) = -2.63, p = .009, d = 0.21$. While the Agent effect is not significant following the rule of thumb of $\alpha = .05$, we consider important to report the results. Participants reported more often negative information about the robot ($mean = 0.10, SD = 0.30$) compared to the human ($mean = 0.05, SD = 0.22$), $t(682) = -1.93, p = .054, d = 0.16$.

    \item[\scalebox{0.8}{$\bullet$}] \textbf{Sanitation}. 
    Results showed a main effect of the Agent on the perception of sanitary issues, $F(1,682) = 8.15, p = .004, \eta^2 = 0.012$.
    More specifically, participants reported more sanitary issues when dealing with the robot ($mean = 0.11, SD = 0.32$) compared to the human ($mean = 0.06, SD = 0.23$), $t(682) = -2.85, p = .004, d = 0.23$.

    \item[\scalebox{0.8}{$\bullet$}] \textbf{Food from Strangers}. 
    Results showed a main effect of the Agent ($F(1,682) = 43.31, p < .001, \eta^2 = 0.052$), the Embodiment ($F(1,682) = 113.69, p < .001, \eta^2 = 0.130$), and their interaction ($F(1,682) = 30.24, p < .001, \eta^2 = 0.034$).
    More specifically, people reported more often refusing food from strangers when the Agent was a Human ($mean = 0.38, SD = 0.49$) compared to a Robot ($mean = 0.11, SD = 0.32$), $t(682) = 6.73, p < .001, d = 0.55$. People also reported more often refusing food from strangers when the agents were screen embodied ($mean = 0.35, SD = 0.48$) compared to when agents were physically embodied ($mean = 0.03, SD = 0.16$), $t(682) = 10.7, p < .001, d = 0.87$. 
    Post-hoc analyses have also been conducted to understand better the Agent $\times$ Embodiment interaction effect.
    All effects were significant except the difference between the physically embodied human and robot, $t(682) = 0.75, p = .877$. People declared more often refusing food from strangers when the human making the offer was screen embodied ($mean = 0.54, SD = 0.02$) compared to physically embodied ($mean = 0.05, SD = 0.04$), $t(682) = 11.31, p < .001, d = 1.32$. Refusing food from strangers was also specified more often for screen embodied human ($mean = 0.54, SD = 0.02$) compared to a screen embodied robot ($mean = 0.17, SD = 0.02$), $t(682) = 10.74, p < .001, d = 1.00$ and also compared to a physically embodied robot ($mean = 0.01, SD = 0.03$), $t(682) = 12.47, p < .001, d = 1.42$.
    People also declared more often refusing food from strangers when a screen embodied robot was making the offer ($mean = 0.17, SD = 0.02$) compared to a physically embodied human ($mean = 0.05, SD = 0.04$) and a physically embodied robot ($mean = 0.01, SD = 0.03$), respectively $t(682) = -2.74, p = .032, d = 0.32$ and $t(682) = 3.69, p = .001, d = 0.42$.

    \item[\scalebox{0.8}{$\bullet$}] \textbf{Etiquette}. 
    There is only a interaction effect between the Offer and the Embodiment on the Etiquette behaviors, $F(1,682) = 14.16, p < .001, \eta^2 = 0.020$.
    Post-hoc analyses have been conducted to understand better this interaction. When agents were screen embodied, results showed an Offer effect associated with more Etiquette behaviors in the Cookie condition ($mean = 0.05, SD = 0.22$)  compared to the Cookie + \$2 condition ($mean = 0.03, SD = 0.16$), $t(682) = 3.65, p = .002$. The Etiquette behaviors were also higher when physically embodied agents made a Cookie + \$2 offer ($mean = 0.08, SD = 0.02$) compared to the same offer made by screen embodied agents ($mean = 0.00, SD = 0.01$), $t(682) = -3.65, p = .002$.
\end{itemize}

\subsection{Explanation and Intentionality}
\begin{figure*}[htbp]
  \centering
  \begin{subfigure}[b]{0.45\textwidth}
    \includegraphics[width=\textwidth]{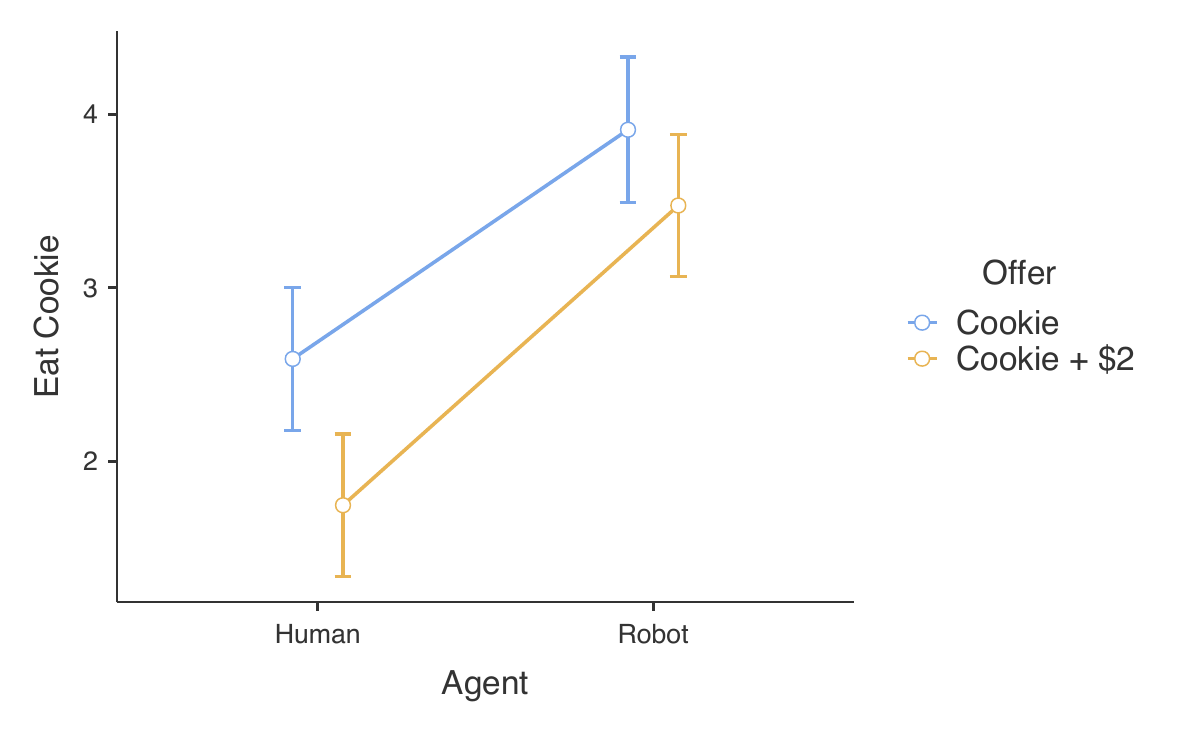}
    \caption{I would eat the cookie.}
    \label{fig:likert_sub1}
  \end{subfigure}
  \hfill
  \begin{subfigure}[b]{0.45\textwidth}
    \includegraphics[width=\textwidth]{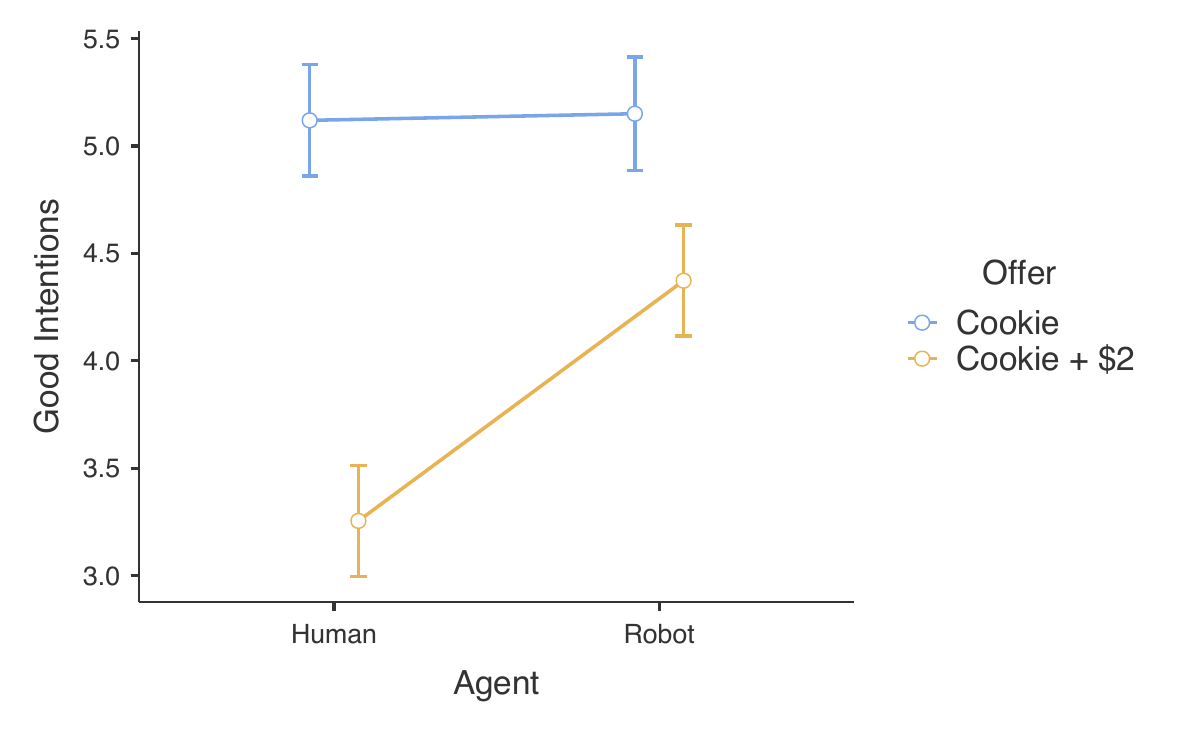}
    \caption{The person/robot had good intentions for offering me the cookie.}
    \label{fig:likert_sub2}
  \end{subfigure}

  \begin{subfigure}[b]{0.45\textwidth}
    \includegraphics[width=\textwidth]{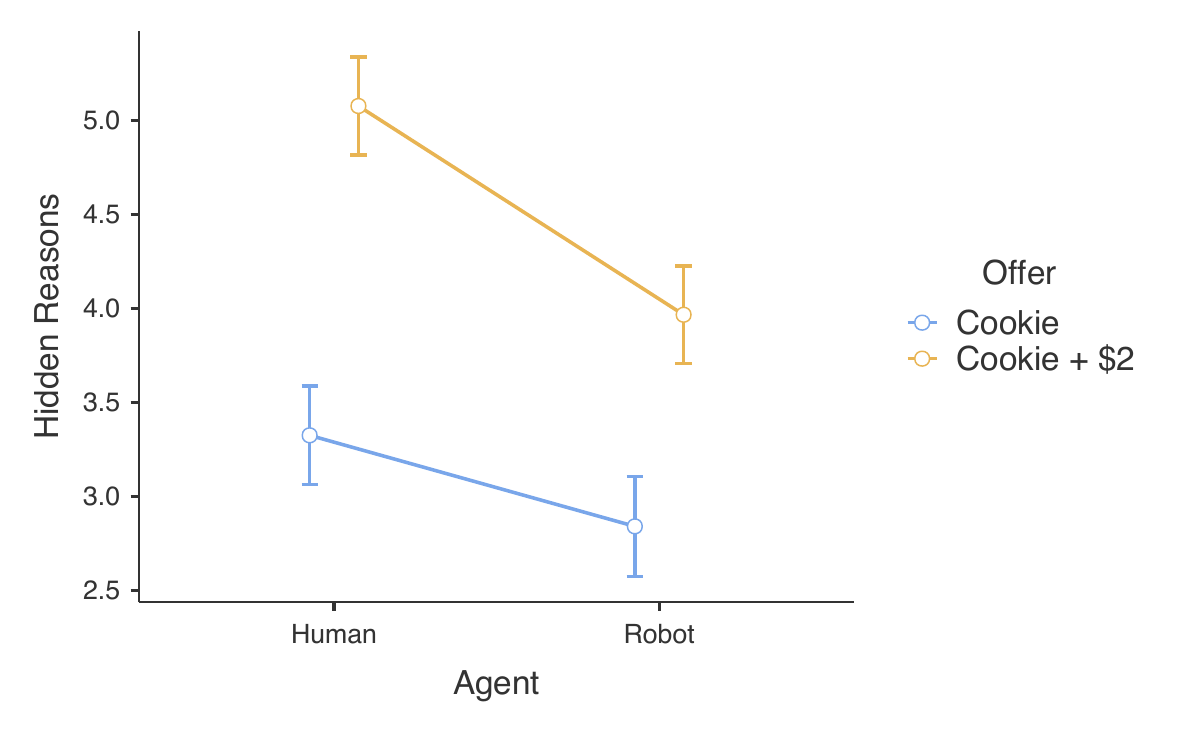}
    \caption{The person/robot had hidden reasons for offering me the cookie.}
    \label{fig:likert_sub3}
  \end{subfigure}
  \hfill
  \begin{subfigure}[b]{0.45\textwidth}
    \includegraphics[width=\textwidth]{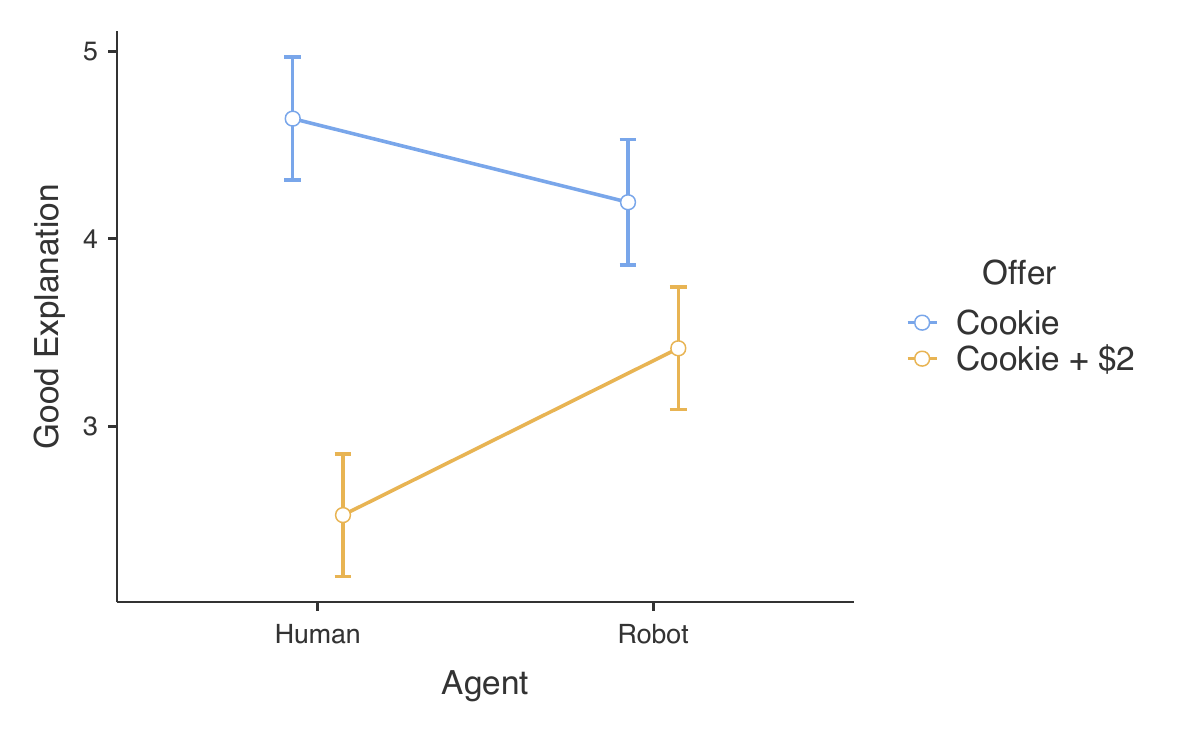}
    \caption{The person/robot gave a good explanation for why he/it offered the cookie.}
    \label{fig:likert_sub4}
  \end{subfigure}

  \caption{Graphs Representing the Distribution of the Likert Scales.}
  \label{fig:likert_scales}
\end{figure*}

Analyses of Variance (ANOVA) have been conducted to see how both factors Agent and Offer influenced each of the 7-point Likert scales. Plots are shown in \autoref{fig:likert_scales}. Note these questions have not been asked to participants who met a physically embodied agent.

When participants have been asked to rate on a 7-point Likert scale the statement ``I would eat the cookie'' (\autoref{fig:likert_sub1}), data analyses showed an Agent and an Offer effect. In other words, participants were more likely to accept the offer made by a robot ($mean = 3.69, SD = 0.15, 95\%CI[3.40, 3.99)]$) compared to a human ($mean = 2.17, SD = 0.148, 95\%CI[1.88, 2.46)]$), $F(1, 462) = 52.75, p < .001, \eta^2 = 0.100$. Participants were also more likely to accept a cookie alone ($mean = 3.25, SD = 0.149, 95\%CI[2.96, 3.54)]$) compared to when it was associated to a monetary outcome ($mean = 2.61, SD = 0.148, 95\%CI[2.32, 2.90)]$), $F(1, 462) = 9.30, p = .002, \eta^2 = 0.0.18$.

When participants were asked to rate the statement ``The person/robot had good intentions for offering me the cookie'' (\autoref{fig:likert_sub2}), ANOVA analyses showed a main effect of the Agent ($F(1, 462) = 18.9, p < .001, \eta^2 = 0.032$) and the Offer ($F(1, 462) = 99.9, p < .001, \eta^2 = 0.167$) as well as an interaction effect between these two factors ($F(1, 462) = 16.9, p < .001, \eta^2 = 0.028$). Post-hoc analyses with Tukey corrections showed that the human offering a cookie was considered as having more good intentions ($mean = 5.12, SD = 0.132, 95\%CI[4.86, 5.38)]$) than the human offering the cookie associated with a monetary outcome ($mean = 3.25, SD = 0.131, 95\%CI[3.00, 3.51)]$), $t(462) = 10.02, p < .001, d = 1.31, 95\%CI[1.04, 1.58]$. The same conclusion can be drawn for the robot offering a cookie ($mean = 5.15, SD = 0.134, 95\%CI[4.89, 5.41)]$) compared to the robot offering a cookie with \$2 ($mean = 4.37, SD = 0.131, 95\%CI[4.11, 4.63)]$), $t(462) = 4.14, p < .001, d = 0.55, 95\%CI[0.28, 0.81]$. While no significant difference was reported between the human and the robot when they both offer a cookie ($t(462) = -0.16, p = .999$), the robot was judged as having more good intentions than the human when they both offered a cookie associated with a monetary outcome ($t(462) = -6.02, p < .001, d = 0.78, 95\%CI[0.52, 1.05]$).

When participants rated the statement ``The person/robot had hidden reasons for offering me the cookie'' (\autoref{fig:likert_sub3}), ANOVA analyses showed a main effect of the Agent ($F(1, 462) = 35.8, p < .001, \eta^2 = 0.060$) and the Offer ($F(1, 462) = 116.58, p < .001, \eta^2 = 0.188$) as well as an interaction effect between these two factors ($F(1, 462) = 5.52, p = .020, \eta^2 = 0.009$). Post-hoc analyses with Tukey corrections showed that a human offering a cookie associated with a monetary outcome ($mean = 5.08, SD = 0.132, 95\%CI[4.82, 5.34)]$) was considered as having more hidden reasons than if it was a simple cookie ($mean = 3.32, SD = 0.133, 95\%CI[3.06, 3.59)]$), $t(462) = -9.34, p < .001, d = 1.22, 95\%CI[0.95, 1.49]$. The same is true for the robot offering a cookie with \$2 ($mean = 3.97, SD = 0.132, 95\%CI[3.71, 4.23)]$) compared to a simple cookie ($mean = 2.84, SD = 0.135, 95\%CI[2.57, 3.11)]$), $t(462) = -5.95, p < .001, d = 0.78, 95\%CI[0.52, 1.05]$. While the difference between both human and robot agents offering a cookie alone is barely not significant ($t(462) = 2.55, p = .053, d = 0.34, 95\%CI[0.08, 0.60]$), the human offering a cookie and \$2 was considered as having more hidden reasons to justify such behavior compared to the robot offering the same offer, $t(462) = 5.93, p < .001, d = 0.77, 95\%CI[0.51, 1.03]$.

When participants rated the statement ``The person/robot gave a good explanation for why he/it offered me the cookie'' (\autoref{fig:likert_sub4}), ANOVA analyses showed a main effect of the Offer ($F(1, 462) = 74.73, p < .001, \eta^2 = 0.135$) and an Agent $\times$ Offer interaction effect ($F(1, 462) = 14.92, p < .001, \eta^2 = 0.029$). No significant effect of the Agent has been reported ($F(1, 462) = 1.75, p = .186$). As a reminder, the explanation provided by the agents was that they were eating with friends and have left-overs. Post-hoc analyses with Tukey corrections showed that the explanation provided by a human offering the cookie alone was considered as a better explanation ($mean = 4.64, SD = 0.167, 95\%CI[4.31, 4.97)]$) compared to when he was offering the cookie associated with a monetary outcome ($mean = 2.53, SD = 0.166, 95\%CI[2.20, 2.85)]$), $t(462) = 8.97, p < .001, d = 1.17, 95\%CI[0.90, 1.44]$. The same is true for the robot offering a cookie alone ($mean = 4.19, SD = 0.170, 95\%CI[3.86, 4.53)]$) compared to a cookie with \$2 ($mean = 3.42, SD = 0.166, 95\%CI[3.09, 3.74)]$), $t(462) = 3.28, p = .006, d = 0.43, 95\%CI[0.17, 0.69]$. While no significant difference was found between both agents when they were both offering a cookie alone ($t(462) = 1.87, p = .241$), offering a cookie associated with a monetary outcome of \$2 influenced the quality of the explanation according to the Agent: a robot offering a cookie and \$2 because it had left-overs after eating with friends was considered as a better explanation compared to the human making the same offer and giving the same explanation, $t(462) = -3.78, p = .001, d = 0.49, 95\%CI[0.24, 0.75]$.

Pearson's correlation analyses have been conducted to explore the relationship between the ``Good intentions'', ``Hidden reasons'', and ``Good explanations'' for both agents separately.

For the \textit{Robot} agent, results showed that the better the explanation was, the better intentions people thought the robot had $(r(229) = .535, p < .001)$ and the hidden reasons the robot had to make an offer $(r(229) = -.376, p < .001)$. The robot having good intentions and hidden reasons were negatively and moderately correlated where the better intentions the robot had, the less hidden reasons to make an offer $(r(229) = -.654, p < .001)$.

For the \textit{Human} agent, results showed that the better the explanation was, the better intentions people thought the robot had $(r(233) = .660, p < .001)$ and the hidden reasons the robot had to make an offer $(r(233) = -.557, p < .001)$. The robot having good intentions and hidden reasons were negatively and strongly correlated where the better intentions the robot had, the less hidden reasons to make an offer $(r(233) = -.776, p < .001)$.

\subsection{Antisocial and Social Behaviors}
While using an online study with screen embodied agents allowed researchers to specifically investigate what they were looking for, using physically embodied agents led people to behave in some ways that were originally intended to be taken into account.

These behaviors are qualitatively discussed in this section and can be divided into social and antisocial behavior.
The latter (i.e., antisocial behaviors) fall into two categories: \textit{ignore} and \textit{abuse}. The former corresponds to actively ignoring the agent (i.e., paying attention to the agent/hearing what the agent is saying but choosing to ignore the agent and/or the interaction), while the latter is divided into two sub-categories: \textit{touching} (i.e., touching the agent) and \textit{stealing} (i.e., taking the agent's offer without the agent having interacted with the person).

\subsubsection{Social Behaviors}
Participants in this section were approached and consented to have their data (i.e., gender) collected for data analysis and publication.
Three persons (one male and two females) waved at the robot when it greeted them. The male returned several times each day to greet the robot and stand in front of it to try to interact with it. However, he had been informed by the experimenter that the robot was not autonomous. Several participants interacted with the robot, trying to initiate a conversation with it (e.g., ``What is your name?'', ``How are you?'', or ``You want me to take a cookie?'' after the robot made the offer). One female even approached the robot to ask directly if she could have a cookie. Another female specified having accepted the cookie from the robot because it was polite.

\subsubsection{Antisocial Behaviors}
Note that participants discussed in this section were not approached by the experimenter after having exhibited antisocial behaviors, except when they have ignored the human agent.

\paragraph{Ignoring Behaviors}
People actively paying attention to the agents but ignoring them (e.g., looking at them, listening and keep walking) were reported as ignoring behaviors. These numbers only included those who looked at the agent after he/it talked and waved to people by looking at them, and/or slow done looking at the robot when it talked and finally continued walking then. Participants with headphones or who did not look at the agent at all when he/it tried to interact with people were not counted as we could not ensure they hard the agent talking. When this behavior happened, the student volunteers were asked to report this on their sheet. Many people ignored the agents when they called. However, people were more likely to ignore the robot (n = 388) than the human (n = 17, including 12 males and 5 females). For this reason, it was challenging for the student volunteers to manage to go towards people in the robot conditions and ask for their gender. For the human conditions, as this was less frequent, the experimenter could ask them directly this information. Chi-square analyses with continuity correction indicated a significant difference between the human and the robot, $\chi^2(1) = 173, p < .001$. The effect size was large, as indicated by Cramer's V = 0.518. It is interesting to note that while some chose to simply ignore the robot, others (two males and one female) chose to run away from the robot when it spoke or waved to them. To do this, one person turned around and took a different route, while most of the people fleeing the robot created a very high safety zone, sometimes bypassing the robot by more than ten meters.

\paragraph{Abusive Behaviors}
No abusive behavior was recorded for the two conditions involving the human agent, whereas the conditions involving the robot led to two types of abuse: stealing and touching the robot. No information regarding these people were collected except their behaviors and presumed gender. Information on cookie theft was collected in order to keep a record of the number of cookies still available for the study and to assure that a sufficient number of cookies was still in the container. Information relating to physical abuse of the robot was collected so that the experimenters could keep track of the risks, and to the extent of prevention, that one of these people might have caused the robot to malfunction.

Twenty-one people (18 males and 3 females presumed) stole cookies from the robot, while no one stole money. Nine people (6 males and 3 females presumed)  uninvited touched the robot. Among these nine persons, eight (6 males and 2 females presumed) went so far as to click the robot's power button repeatedly. When people touched or stole from the robot, Nao said ``Do not touch me'' and ``Do not take a cookie without my permission'' respectively. While two people quickly walked away after hearing this message, most of the others felt amused and touched the robot more to the point that experimenters had to go to tell them to stop touching the robot. Moreover, data visually showed that males were more keen to abuse the robot compared to females, either by touching or stealing from the robot. 

\paragraph{Flight and Personal Space}
Some people, after the robot attempted to interact with them, avoided the robot by walking at sometimes more than 10 meters from it or make a U-turn.

\section{Discussion}
\subsection{Phantom Costs}
Phantom costs are based on the idea that people do not aim to only maximize potential gains \citep{vonasch2022} but interact with other humans considering them as social agents. Results show that phantom costs do occur when a human is interacting with a human (RQ1) and a robot (RQ2). This means that when people are offered something (here, a cookie) with or without a monetary outcome of \$2, people tend to less often accept the offer when it comes with extra money. This shows that people naturally consider humans as social agents but do also with social robots (see \citep{Nass1994, ReevesNass1996}.

The current research shows that people imagine phantom costs when they consider explanations provided as insufficient. Therefore, money causes suspicion and people start imagining hidden reasons to explain this extra generosity, as reported by participants when they were asked to justify their decisions. The perception of phantom costs leading to a money backfire effect (i.e., people refuse an offer more often when associated with extra money) is consistent with previous research conducted in human-human interaction \citep{vonasch2022}.

Perception of phantom costs in overly generous offers without explicit explanations for why the agent is doing this is consistent with the results obtained in the screen embodied version of the study, as questions regarding the quality of the explanation provided by the agents and the reasons why they would have made such offer were only conducted for screen embodied agents. Results show that for both agents, the more people would have eaten the cookie, the more they considered the agent's explanation to be good, and imagined the agent to have good intentions and fewer hidden reasons. These results support the Heuristic of Sufficient Explanation model \citep{HOSEVonasch} and are consistent with previous work stating that people do attribute robots intentions \citep{Marchesi2019, Thellman2017}.

As social robots are expected to be more and more present in our daily-lives and are already used in healthcare services \citep{Ragno2023} but also to assist and serve humans \citep{Xu2024}, it is important that people correctly understand robot's intentions and consider them as sufficient for people to not consider them as threat and rather trust them \citep{Mou2020}. A way to do so is by attributing them mental states to understand and predict their behaviors \citep{Dennett1987}. As phantom costs are found in both human and robot conditions, sufficient explanations provided by humans have also to be provided by the robot to understand its actions and intentions, especially when they are not understandable by themselves (see \citep{Angelov2021, Aronson2018, Malle2011, Phillips2020, Skow2023, HOSEVonasch}).


\subsection{Agent: Human and Robot}
An interesting part of the analyses corresponds to the social interaction involving a human and a robot. Overall, people accepted more often an offer made by a robot compared to a human. However, such difference is only found when agents were screen embodied, regardless what the offer was, but not when they were physically embodied.
The Agent and the Offer variables independently contribute to the likelihood of participants of accepting or refusing the offer (RQ3). This means that the effect of the Agent on the choice people make (i.e., accept or refuse the offer) does not change depending on the level of the Offer.

The reason of these results might be related to the expectations of social cues from the human and their absence in the screen embodied condition. A physical interaction involves many social and non-verbal cues (e.g., gaze, posture) whereas our screen embodied agents were limited in exhibiting such cues. We might postulate that a dissonance occurs with the screen embodied agent (as a picture) and what participants would have expected from a human compared to a robot (i.e., more social cues from humans). If so, participants' expectations about the screen embodied human were not met whereas no such high expectations were intended to be present for the robot. On the contrary, when agents were physically embodied, both the human and the robot might have met people's expectations regarding their social behaviors. This is relevant with \citet{Thellman16} who specified that it is not the physical presence that matters but the social presence. We consider that both matter but that they should meet expectations people have towards them which also depends on the physical presence. \citet{Xu2015} found a significant difference in terms of social presence in favor of a live interaction with a robot compared to a picture associated with a textual scenario. It remains possible to think that participants did not feel social presence of the robot even when it was physically embodied.
Another reason may be that people consider less risks when interacting with the physically embodied human because of the context (a person of student age, located at the University) while the context used in the screen embodied study was simply to go for a walk, without specifying a specific location. In the former case, it may appear easier to trust a student at the University compared to a complete stranger when we are going for a walk.

\subsection{Embodiment: Screen and Physical Agents}
Focusing on the Embodiment factor, people were more likely to accept an offer (a cookie alone or a cookie associated with \$2) made by a physically embodied human compared to a screen embodied human. No such difference was found when the agent was a robot for both screen and physical embodiment levels. 
Phantom costs occurred in both embodiment levels though, effect sizes of the Offer effect (Cookie - (Cookie + \$2)) were larger when agents were physically embodied compared to screen embodied. This supports the results showing more phantom costs described by participants in the screen conditions. While \citet{vonasch2022} only used text-written scenario, \citet{xu2012} showed that a text associated with a photo provides pretty good results when assessing attitudes towards robots. The stimuli of this study were pictures with text-written scenario and data analyses highlight phantom costs effects.
While no study, to our knowledge, has previously explored this human-robot interaction in the literature, we can only compare the HHI results. These results are consistent with \citet{vonasch2022}. In their study conducted in person, \citet{vonasch2022} found that more people accepted the cookie alone ($M = 39.4\%$) compared to when the offer was associated with money ($M = 20.5\%$). Our results show stronger pieces of evidence as, for the human conditions, 64.8\% of participants have accepted to eat the cookie compared to only 24.1\% when the offer was associated with a monetary outcome of \$2. These results have to be put into perspective by the geographical location of the participants: while \citet{vonasch2022} conducted their study in the USA (specifically, in Chapel Hill, North Carolina), the present study was conducted in New Zealand. The higher rate of acceptance in both human conditions may reflect the higher level of interpersonal trust in New Zealand than in the USA as shown in The Social Report 2016 – Te pūrongo oranga tangata\footnote{\url{https://socialreport.msd.govt.nz/social-connectedness/trust-in-others.html}}; New Zealand is one of the safest countries in the world (4th position) while the USA are at the 131st position according to the 2023 Global Peace Index\footnote{\url{https://www.visionofhumanity.org/maps/}}.

Combining the variables Agent, Offer, and Embodiment does not exhibit a statistically significant impact on the likelihood of accepting the offer (RQ4). 
In the same vein, combining the Embodiment and Offer variables does not influence differently the likelihood of accepting the offer (RQ6) showing that these two variables influence independently participants' choice regardless the agent embodiment.

\subsection{Decision Criteria}
Participants were asked to justify their decision to accept or refuse the offer made by the agents to understand better decision-making processes underpinning their choices. Participants use several reasons to explain their choice of accepting or refusing the offer made by the agent (RQ5).

Reasons to accept the offer were to (i) be excited about the offer (the cookie and/or the money), (ii) perceiving the agent as good (e.g., trust, the agent perceived as trustful or kind), (iii) being hungry, or even (iv) the etiquette (accepting the offer to not look rude). reasons to refuse the offer were  (i) to perceive phantom costs (i.e., finding something suspicious), but also (ii) not being hungry, (iii) perceiving the agent as bad (e.g., not trustful, have bad intentions), (iv) sanitation issues or simply (v) following the rule ``no food from strangers.''


Our results show that phantom costs perception, excitement about the offer, and refusing food from strangers are major predictors of participants' decisions. While the latter (refusing food from strangers) is mainly food in the screen conditions, phantom costs perceptions and excitement are found in both screen and physically embodied agents. These results support our hypotheses and previous research \citep{vonasch2022} stating that phantom costs and being excited about the offer are major predictors explaining why people make their decisions to accept or refuse an offer. People perceive the agent more positively when the offer is not associated with a monetary outcome compared to when it is, regardless whether it is a human or a robot, and the reverse is also true. This shows that if an offer is too generous, people tend to attribute bad intentions to the agent and trust it less.
The differences between human and robot agents in terms of participants' negative perceptions are barely not significant ($p = .054$), but go in the direction of people having more negative perceptions of the robot compared to the human agent. This can be explained due to the eeriness people may feel when interacting with a robot for the first time (see \citet{Belpaeme2020, Mori2012, Paetzel2020}).

While the HOSE model expects people to try to find hidden reasons for events for which they judge explanations to be insufficient, many participants judged the interaction with Nao inconsistent with their knowledge and expectations as robots do not eat and do not have friends. This inconsistency led people to perceive phantom costs associated with the offer, but also had other consequences. This absence of relevant information from the robot sometimes mitigated the robot autonomy perceived by participants. This brought participants to consider that cookies were made by someone else and people did not trust the human behind all of this.
Making a choice as a function on how we perceive the agent making the offer is supported by the literature on trust. \citet{Mou2020} showed that attributing a mind to the robot pushes people to trust it more. This seems to go against our results associated with the decision-criteria participants specified having used.

It is worth specifying that participants qualitatively described more often positive and negative perceptions of the robot than the human, this is likely due to the the likelihood of participants to interact for the first time with a robot and show curiosity or fear and thus increase the likelihood of justifying their choice by describing how they perceived the robot. This perception supports the HOSE model \citep{vonasch2022} as participants perceived more negatively agents when the offer was associated with a monetary outcome while no sufficient explanation was provided, and more positive perceptions of agents when they just offered a cookie as explanations were more sufficient.
Despite the barely not significant difference in terms of bad perceptions of the agent, it is possible to think that people might trust robots less than humans because they are more easily attributing intentions to humans and these intentions can predict/understand better their behaviors. Moreover, people seem to trust more easily Nao by its robotic nature (e.g., ``It is a robot. I don't think it would poison me.'', or ``I don't think a robot would lie to me or have ill intentions.''). Some people have expectations about robots such as respecting Asimov's laws by being harmless to humans.

Etiquette behaviors (e.g., accepting the offer because the agent made it, it would be impolite to refuse) are used by participants for the cookie condition when agents are screen embodied and more in the cookie + \$2 condition when agents were physically embodied. This difference can be explained by differences in the agent embodiment and how people perceive risks. We suggest two reasons for this. First, in a hypothetical scenario (i.e., online setting with the screen embodied agents), people might not perceive risks the same way or will not attribute a same level of social norms towards humans and robot. This relates to social trust which is defined as the expectation that people will behave well without attempting to harm others (see \citet{Freitag2013}). Participants may simply not know robot's intentions as they certainly never interacted with one before. Participants may consider it more disrespectful to ignore a human and refuse the offer he made compared to a inanimate robot. This result may goes
Second, participants may have been excited about the fact of being paid to eat in the physically embodied agent conditions but do not specify this to the experimenter when asked to explain their behaviors.

\subsubsection{Behaviors Towards Physically Embodied Agents}
Because in the physical embodiment condition, participants were unaware that a study was being conducted, it provided an opportunity to observe how people might actually respond to a robot in the wild, rather than in tightly controlled laboratory settings where everyone knows a study is being conducted. Some people abused the robot, others were frightened by it, and others ignored it, as well as many people responding to it as if it were a social agent offering them something. 

Several people abused the robot, including stealing from it and touching it aggressively. These behaviors were more common among males than females. Prior research has shown similar antisocial behaviors toward robots \citep{Brscic2015, Ku2018, Nomura2016, Smith2018}. However, most studies of abusive behavior towards robots have shown that such behaviors are common among children \citep{Brscic2015, Nomura2016}. Our study shows that they can also be found in young adults of university student age. Moreover, the robot asked its abusers to stop abusing it, and people's responses were mixed. After the robot asked people not to steal or touch it, some people stopped, but others laughed and continued to do so, sometimes until the experimenter had to intervene. Thus, like with preventing abuse in humans, simply asking the abuser to stop is not always sufficient. Asking may even encourage the abuse to continue if the abuser is rewarded by thinking the robot's protestations are funny \citet{Ku2018}. Therefore, Nao's request could be seen as an encouragement to further abuse the robot. Further studies seem necessary to find the best way for the robot to discourage abusive behavior and to protect itself. These abusive behaviors did not occur towards humans and may therefore reflect the way people perceive the robot. We cannot know which of the featural differences between humans and robots caused people to abuse the robot but not the human. One possibility is the submissiveness of the Nao robot \citep{Rosenthal2014}, but it might also be its height, its novelty in the social environment, or some other feature.

Our results have to be put in perspective with potential participants who chose to simply ignore the agents and so neither accepted or refused the offer. Very few participants ignored the human (n = 17), but hundreds (n = 388) ignored the robot. There are several potential reasons for this. One is because the robot's voice is not human, people may not have heard it, or may not have recognized that it was talking to them specifically. Another reason is people may have felt uncomfortable having an interaction with a robot outside of a laboratory setting--they might not have known what to do or how to talk with a robot. The social norms are unclear. They may have also worried about being disrespectful to a human asking them a question, but felt not such worries about disrespecting a robot. Or they may have assumed the robot was not as technologically advanced as it appeared, so justified ignoring it \citet{Schneider2022}.

\subsection{Limitations}
Some limitations have to be discussed.
The first one corresponds to the generalisability of the results to human-robot interaction. We have used a specific robot (Nao) which has a certain degree of human-likeness, we expect our results to be influenced by the degree of human-likeness of the robot people will perceive. We cannot assure such conclusions are still true, at least to the same extent, with non human-like robots. Also, while \citet{vonasch2022} only recruited adults sitting alone outside, we have attempted this but a huge proportion of participants who approached the robot were in group as most of the people alone just ignored the robot. We cannot exclude potential group biases in these robot conditions, but results are similar to those of the physically human conditions and phantom costs did occur. A strength of this is that he shows better how people behave and perceive robots when alone and in groups. This difficulty to recruit people alone was not the case for participants who approached the human. This may be due to discomfort alone people felt to be called out by a robot for which they did not know its intentions.

Then, our online study was conducted with Americans and the in person study was conducted in New Zealand. This choice was made for pragmatic reasons: our university is in New Zealand, but Prolific's sample of New Zealanders was too small to efficiently generate a suitable sample size, so we used Americans instead. This also allowed us to test this effect across both cultures, increasing generalisability of the results, but because we used different methods (in person versus online) for the two samples, it does not allow us to directly compare the two cultures in how people respond in these situations. People in both cultures showed the effect, but it is not possible to know whether the effect might be larger in one culture than the other. 

Concerning the screen embodied agents version, one might criticize that authors have chosen to use pictures over video and/or text-written interactions. \citet{Xu2015} indicated that the chosen media should depend on the study purpose. The authors indicated that texts and pictures are low in both interactivity and presentation fidelity. While video-based scenario shows several aspect that cannot be shown with a picture (e.g., dynamic of the conversation), we showed that the use of pictures is enough to trigger phantom costs. Moreover, the effect also occurred in real life interactions with a robot, suggesting the phantom costs effect is robust to both real interactions and simulated text and photo based, and in both HHI and HRI. 

In the same idea of comparing the embodiment, these studies were conducted either online or in-person, which provided methodological tradeoffs. For example, online participants were more likely to explain the reasoning behind their choices (because that format allowed us to easily ask them to provide reasons), whereas in-person participants rarely did (because we wanted the interactions to be as realistic as possible). However, studying real in-person behavior has substantial advantages in terms of how realistic people's responses are, and how likely they are to generalize to real world behavior.


Concerning the physically embodied agents part, as social robots such as Nao are still rarely present in people's life, participants who were in the robot conditions may have interacted with a robot for the first time in their lives. This first exposure might have implied a novelty effect (see \citet{Belpaeme2020} that could have influenced the likelihood of accepting the offer. Some participants were visibly excited by the idea that a robot was talking to them, and they might therefore have been more likely to accept its offer, due to the novelty and excitement of the situation. However, others were visibly apprehensive and cautious about talking to the robot, presumably because it was an unfamiliar situation for them. This might have made them less likely to accept its offer. Overall, people were no more willing to accept the offer from a robot than from a human, suggesting that neither the excited nor the apprehensive participants predominated our sample enough to bias it in one direction or the other.

In the original study \citet{vonasch2022}, the agent was a human who walked toward participants to make the offer. However, in this study, the robot could not walk to approach people, so we had to design a different way for it to engage participants: It called out to them instead. To be consistent across conditions, in the human condition, the human also called out to participants without walking toward them. This difference from the original paradigm may have affected our results, though it is not possible to know which type of engaging participants would make them more likely to accept versus reject the offer. 

Finally, a large portion of people who the robot called out to ignored the robot. Thus, we do not know whether they would have accepted its offer had they been offered it. The proportion of people ignoring the robot was substantially more than those who ignored the human, so this likely was due to some characteristic of those people. It is possible that if the people who ignored the robot had been able to participate in the study, the proportion of people accepting the offer might have been different. There is no way to know, and no way to fix this issue without potentially undermining the ecological validity of the situation, which is one of the major strengths of this research. 

The study had a very strong ecological validity for three main reasons. First, participants were not aware beforehand that they were participating in a study when they started to interact with the agent. Second, running the study with physically embodied agents, especially with the robot, allowed the experiments to report observational data (i.e., social and anti social behaviors). Finally, even if the robot conditions were conducted using the Wizard-of-Oz method, almost no participants reported the robot's non-autonomy in their comments.

\section{Conclusion and Future Work}
This study explored whether phantom costs can be found in an interaction involving a human and a robot, where agents were either physically or screen embodied. Phantom costs occur when people are suspicious and search for hidden reasons for an event when explanations seem insufficient. Such behaviors are evidence that people are social agents and not fully rational.
To explore this, a human or a robot offered people either a cookie or a cookie associated with a monetary outcome of \$2. The latter is an overly generous offer with no known reason to explain it. Whether people accepted the offer was the dependent variable. While rational behaviors associated with conventional economic models would lead people to accept offers that seem more advantageous, the Heuristic of Sufficient Explanation (HOSE) economic decision-making model shows that people will engage themselves in social cognition by attempting to understand the intentions of the person making such generous offers.

Results show that phantom costs do occur in HHI and HRI in both embodiment levels (screen and physically embodied agents), more frequent with screen embodied agents though. Also, people tend to accept more often offers made by a robot compared to a human.
This shows that the HOSE model and the concept of phantom costs can be extended to human-robot interaction.

This provides information on how people interact and perceive social robots compared to humans.

This indicates that people consider both humans and robots as social agents and behave socially but irrationally with both. This even if money has no real value and meaning for the robot. 

This provides information on how people interact and perceive social robots compared to humans. People are able to imagine hidden reasons from robots behaviors by attributing them intentions.
Thus, we show that it is possible to attribute positive and negative intentions to robots and that they influence human decision-making processes.

Such information is important to be known as social robots are expected to be used as assistants and companions by helping humans for several tasks in different contexts such as healthcare services. In such contexts, robots it is important to ensure that robots will provide sufficient explanation/reasons justifying their behaviors to ensure people will accept behaviors and help from them. This shows the importance of transparency by adequately justifying robot's behaviors to maintain trust and acceptance from people. Also, this study joins several scholars' side stating that using physically embodied agents for HRI research influences behaviors differently than screen embodied agents but also that both embodiment levels provide significant effects of the variables of interest.

Further research could explore the impact of physical anthropomorphism on the perception of phantom costs and decision-making processes in HRI.
This could also be done using different type of population and experts in robotics to see the impact of the novelty effect on human's behaviors.
We could explore to what extent people can show irrational behaviors towards robots using different types of offers and explanations provided by the robot, which could have more or less meaning for a robot. It might be also interesting to link this better to the concept of intentionality and see how attributing intentions and autonomy to the robot influences the perception of suspicious behaviors (i.e., phantom costs) from the robot.

\section*{Declarations}
\subsection*{Author contributions}
Conceptualization: BL, CB, AV;
Methodology: BL, CB, AV;
Formal analysis and investigation: BL;
Writing - original draft preparation: BL;
Writing - review and editing: CB, AV;
Resources: CB;
Supervision: CB, AV.

\subsection*{Acknowledgements}
We would like to thank Elena Moltchanova for her support on the power analysis, and Kimberley Penrose and Luke Goleman for their help in recruiting participants and coding the data.

\subsection*{Data availability}
All data (raw and data analysis) will be made available in a repository upon acceptance.

\subsection*{Conflict of interest}
The authors have no relevant financial or non-financial interests to disclose.

\subsection*{Funding}
No funds, grants, or other support was received.

\subsection*{Compliance with Ethical Standards}
This study was approved by the University of Canterbury Human Ethics Committee (HREC 2023/93/LR-PS). The participants provided their verbal or written informed consent to participate in this study.

\begin{appendices}
\appendix
\section{Data Analyses Including all Data}
\label{appendix:a}
This Appendix corresponds to the analyses including the data described in the main data analyses of the paper as well as the previous data with screen embodied agents we considered they may influence the results due to their limitations.

\subsection{New Participants}
\begin{figure*}
\begin{center}
\includegraphics[width=\linewidth]{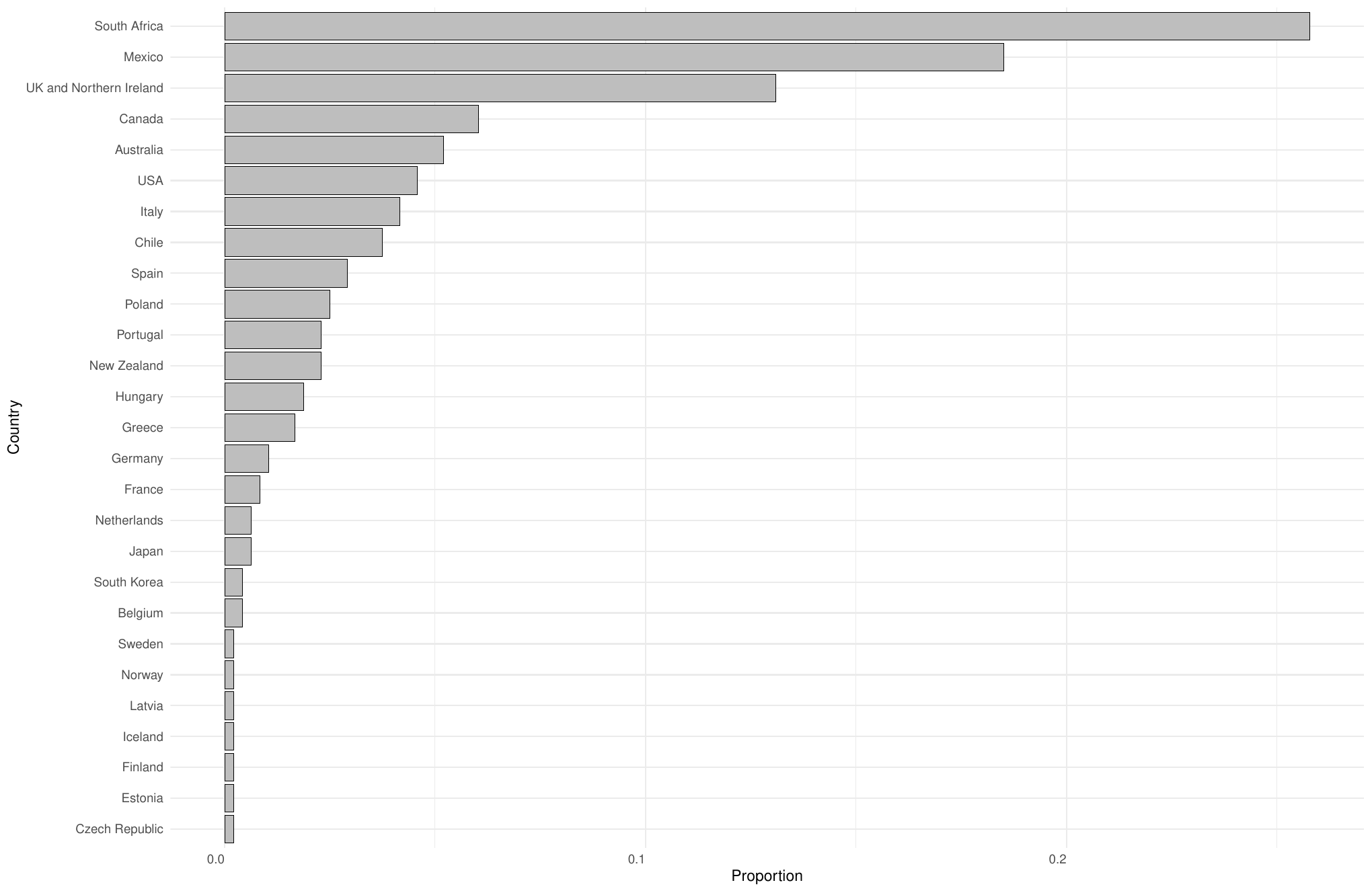}
\end{center}
\caption{Country Distribution of the Participants of the Screen Embodiment Part.}
\label{fig:country}
\end{figure*}

Without describing previous participants of the main paper, 485 participants were recruited via the Prolific platform to take part in this online study. The only inclusion criterion was to be fluent in English. Among the 485 participants, 4 participants were automatically not included in our dataset by Prolific (timed-out), 10 participants were removed from the data analyses because they did not meet the attention check justification (n = 2) or had dietary restrictions (n = 8) leading to a final sample size of 471 participants for this online study. The gender distribution of this final sample (N = 471) was as follows: 235 were males, 222 were females, and 14 announced having a different gender. Their age ranged from 20 to 75 ($Median = 29,M = 32.41, SD = 10.94$). While all participants claimed to be fluent in English, the geographical distribution was broad with 27 different countries (see \autoref{fig:country}). Most participants were from South Africa (about 26\%), followed by Mexico (about 19\%), UK and Northern Ireland (about 12\%).

\subsection{Results}
Logistic regression analyses have been conducted using the R software to examine the relationship between the Agent, the Embodiment, and the Offer the Agents made, on the probability of Rejecting (0) or Accepting (1) the Offer. The Agent was either a Human (0) or a Robot (1). They were either screen (0) or physically (1) embodied. The Offer was either a Cookie (0) or a Cookie + \$2 (1). The equation is shown in \autoref{eq:overall_model_all_data} and the coefficients table in \autoref{tab:CoefficientsOverallModel_all_data}. Main effects are directly obtained from \autoref{tab:CoefficientsOverallModel_all_data}. The interactions are explored using the \textit{emmeans} package.

\begin{figure*}
\begin{equation}
\begin{aligned}
\text{logit}(\mathbb{P}(\text{Choice} = 1)) = & \beta_0 + \beta_1 \cdot \text{Agent} + \beta_2 \cdot \text{Offer} + \beta_3 \cdot \text{Embodiment} \notag \\
&\quad + \beta_4 \cdot (\text{Agent} \cdot \text{Offer}) + \beta_5 \cdot (\text{Agent} \cdot \text{Embodiment}) + \beta_6 \cdot (\text{Offer} \cdot \text{Embodiment}) \notag \\
&\quad + \beta_7 \cdot (\text{Agent} \cdot \text{Offer} \cdot \text{Embodiment})
\label{eq:overall_model_all_data}
\end{aligned}
\end{equation}    
\end{figure*}

\begin{table*}[ht]
\caption{Coefficients of the Three-Way Interactions Logistic Regression Model.}
\begin{center}
        \begin{tabular}{lcccr}
        \toprule
          Effect & Estimate & Standard Error & z value & $Pr(>|z|)$ \\ \midrule
         (Intercept) & -1.0872 & 0.1505 & -7.222 & $< .001^*$ \\
         Agent & 1.0699 & 0.2001 & 5.346 & $< .001^*$ \\
         Offer & -0.6657 & 0.2370 & -2.808 & .005$^*$ \\
         Embodiment & 1.6982 & 0.3223 & 5.269 & $< .001^*$ \\
         Agent $\times$ Offer & 0.1813 & 0.3027 & 0.599 & .549\phantom{*} \\
         Agent $\times$ Embodiment & -1.2454 & 0.4429 & -2.812 & .005$^*$ \\
         Offer $\times$ Embodiment & -1.0939 & 0.4886 & -2.239 & .025$^*$ \\
         Agent $\times$ Offer $\times$ Embodiment & 0.2956 & 0.6545 & 0.452 & .651\phantom{*} \\
         \bottomrule
    \end{tabular}
    \end{center}
\footnotesize{Note: *: $p < .05$}
\label{tab:CoefficientsOverallModel_all_data}
\end{table*}

\subsubsection{Phantom Costs}
The odds of accepting the offer were 1.94 times higher ($CI 95\% [1.23, 3.12]$) when the offer was a cookie compared to when it was a cookie associated with \$2. For each Agent $\times$ Embodiment, the odds of accepting the Offer were higher when the offer was a cookie compared to a cookie associated with \$2. The odds ratio (OR = Cookie/(Cookie + \$2)) of each pairwise comparison were the followings: 1.95 for the screen embodied human, 1.62 for the screen embodied robot, 5.81 for the physically embodied human, and 3.61 for the physically embodied robot.

\subsubsection{Agent: Human and Robot}
The odds of accepting the offer were 2.91 times higher ($CI 95\% [1.98, 4.33]$) when the Robot was making the offer compared to when the Human was making the offer. While no difference has been found between the Human and the Robot when they were physically embodied, neither for the cookie (p = .657) nor the cookie + \$2 condition ($p = .478$), the odds of accepting the offer were higher for the screen embodied robot compared to the screen embodied human when the offer was a cookie (OR = 2.91, $p < .001$) but also a cookie + \$2 (OR = 3.50, $p < .001$).

\subsubsection{Embodiment: Screen and Physical}
The odds of accepting the offer were 5.46 times higher ($CI 95\% [2.94, 10.45]$) when agents were physically embodied compared to screen embodied. Only the embodiment of the human influenced the odds of accepting the offer. The odds of accepting the offer were 5.46 times higher when the human was physically embodied compared to screen embodied ($p < .001$). The only difference in terms of significance with the main data analyses of this paper is that here, we have a significant Offer $\times$ Embodiment interaction effect ($p = .025$) associated with the absence of a previous significant result: the odds of accepting the cookie + \$2 did not significantly differ between both levels of the human embodiment ($p = 0.010$).

\subsubsection{Gender Analyses}
Exploratory analyses have been conducted with all the data to explore potential gender differences for both embodiment levels on the probability of accepting the offer. To do so, ANOVA were conducted and did not show any significant effect of the Gender ($F(2,718) = 0.59, p = .557$).

\section{Wizard-of-Oz Script}
\label{appendix:b}
\paragraph{Overall Preparation:}
\begin{enumerate}
    \item Take all the material in the room.
    \item Go to one of the locations, close to a power socket.
    \item Turn on the laptop, the WiFi router, and the robot. The robot should be charging every time it is on.
    \item Connect the laptop to the WiFi router via the Ethernet cable. Verify it is connected to NaoNetwork. If not, restart.
    \item Launch Choregraph 2.1.4.
    \item Connect Nao to Choregraph: on Choregraph, click on ``Connection'' and ``Connect to'', write the IP address the robot provides.
    \item Check the robot is connected, and that you see the cameras of the robot on Choregraph.
    \item Deactivate ``Autonomous Life'' if still on: press the heart symbol above the robot view. The heart should be empty.
    \item Click on ``Run''. If not working, restart Choregraph (also the robot if still not working).
    \item Run ``Stand Up.''
    \item Run ``Eye LEDs.''
    \item Run ``Ext. R arm'' to extend the right arm.
    \item Place the stick in the right hand of the robot. Attach the box in the position indicated by the Velcro fasteners. Check it is well attached.
    \item Run ``Init'' to position Nao in its initial position.
    \item Add three cookies in the container. If running ``Cookie Dollars'' condition, attach the \$2 coin on the back of the robot’s left hand.
    \item Check the sound is high enough for people to hear it.
\end{enumerate}

\paragraph{Participant Arriving:}
\begin{enumerate}
    \item Move the head of the robot (using the boxes on the screen or moving manually the head clicking on its head on Choregraph). Continue to follow the participant during the interaction.
    \item When someone is close to the robot, run ``Hi'', ``Hey'', or ``Hello.''
\end{enumerate}

\paragraph{If Cookie Condition:}
\begin{enumerate}
    \item Run ``Cookie.'' If the participant does not understand, run ``Repeat Cookie'' (up to two times).
    \item If the person understood, go to the next point. If the person did not understand: approach them, thank them + debriefing + demographic data (specify in the experimenter' notes that this participant did not follow the protocol).
    \item If the participant does not take the cookie: Run ``Not interested'' to end the interaction.
    \item If the participant takes the cookie Run ``Thanks'' to end the interaction.
    \item Approach the participant. Thank them + debriefing + ask for demographic data. If the experimenter has any comment, write in the experimenter's note column.
\end{enumerate}

\paragraph{If Cookie + \$2 Condition:}
\begin{enumerate}
    \item Run ``Cookie Dollars.'' If the participant does not understand, run ``Repeat Cookie Dollars'' (up to two times).
    \item If the person understood, go to the next point. If the person did not understand: approach them, thank them + debriefing + demographic data (specify in the experimenter' notes that this participant did not follow the protocol).
    \item If the participant does not take the cookie: Run ``Not interested'' to end the interaction.
    \item If the person takes the coin, it should automatically say ``Thank you'', if not: run it (clicking on the ``Cookie Dollars'' box, and the output of ``Tactile L. Hand.''
    \item Approach the participant. Thank them + debriefing + ask for demographic data. If the experimenter has any comment, write in the experimenter's note column.
\end{enumerate}

\paragraph{After the Interaction:}
\begin{enumerate}
    \item Run ``Init'' to put the robot in its initial position.
    \item Add a cookie (and a coin if ``Cookie Dollars'')  for the next participant.
\end{enumerate}

\paragraph{Some Extra:}
Some boxes are available to make the interaction more natural according to what the participant says such as ``Yes'', ``No'', ``I did not hear'' (if the experimenter did not hear what the person said to the robot), ``Thanks'', or ``Not interested.''

\end{appendices}

\bibliographystyle{plainnat}
\bibliography{sample}

\end{document}